\documentstyle[11pt]{article}

\newlength{\bredde}
\def\slash#1{\settowidth{\bredde}{$#1$}\ifmmode\,\raisebox{.15ex}{/}
\hspace*{-\bredde} #1\else$\,\raisebox{.15ex}{/}\hspace*{-\bredde} #1$\fi}
\newcommand{\mat}{\left ( \begin{array}{cc}}
\newcommand{\emat}{\end{array} \right )}
\newcommand{\matt}{\left ( \begin{array}{ccc}}
\newcommand{\ematt}{\end{array} \right )}
\newcommand{\matf}{\left ( \begin{array}{cccc}}
\newcommand{\ematf}{\end{array} \right )}
\newcommand{\vect}{\left ( \begin{array}{c}}
\newcommand{\evect}{\end{array} \right )}

\newcommand{\be}{\begin{eqnarray}}
\newcommand{\ee}{\end{eqnarray}}
\newcommand{\beq}{\begin{equation}}
\newcommand{\eeq}{\end{equation}}
\newcommand{\ba}{\begin{array}{ccc}}
\newcommand{\ea}{\end{array}}
\newcommand{\nn}{\nonumber}
\newcommand{\noi}{\vspace{6pt}}
\newcommand{\lG}{\raise.3ex\hbox{$\stackrel{\leftarrow}{G}$}}
\newcommand{\lU}{\raise.3ex\hbox{$\stackrel{\leftarrow}{U}$}}
\newcommand{\lP}{\raise.3ex\hbox{$\stackrel{\leftarrow}{{\cal P}}$}}
\newcommand{\leta}{\raise.3ex\hbox{$\stackrel{\leftarrow}{\eta}$}}
\newcommand{\lOmega}{\raise.3ex\hbox{$\stackrel{\leftarrow}{\Omega}$}}
\newcommand{\ldr}{\raise.3ex\hbox{$\stackrel{\leftarrow}{\delta^r}$}}

\def\beqn{\begin{eqnarray}}
\def\eeqn{\end{eqnarray}}

\def\gtwid{\raise.3ex\hbox{$>$\kern-.75em\lower1ex\hbox{$\sim$}}}
\def\ltwid{\raise.3ex\hbox{$<$\kern-.75em\lower1ex\hbox{$\sim$}}}

\def\r{\ref}
\def\sign{\mbox{sign}}

\def\la{\lambda}

\def\om{\omega}

\def\matA{{\mbox{\bf A}}}
\def\matB{{\mbox{\bf B}}}
\def\matC{{\mbox{\bf C}}}
\def\matD{{\mbox{\bf D}}}
\def\matS{{\mbox{\bf S}}}

\begin{document}
\topmargin -1.4cm
\oddsidemargin -0.8cm
\evensidemargin -0.8cm
\title{\Large{{\bf QCD$_3$ and the Replica Method}}}

\vspace{1.5cm}

\author{~\\{\sc G. Akemann $^a$,
D. Dalmazi $^b$\footnote{On leave from UNESP (Guaratinguet\'{a}-Brazil)},
P.H. Damgaard $^c$} and
{\sc J.J.M. Verbaarschot $^b$}
\\~\\
$^a$ Max-Planck-Institut f\"ur Kernphysik\\
Saupfercheckweg 1\\D-69117 Heidelberg\\Germany
\\~\\
$^b$ Dept. of Physics and Astronomy, SUNY\\Stony Brook,
N.Y. 11794\\U.S.A.
\\~\\
$^c$ The Niels Bohr Institute\\ Blegdamsvej 17\\ DK-2100 Copenhagen {\O}\\
Denmark
}
\date{}
\maketitle
\vfill

\begin{abstract}
Using the replica method, we analyze the mass dependence
of the QCD$_3$ partition function in a parameter range where the leading
contribution is from the zero momentum Goldstone fields.
Three complementary approaches are
considered in this article. First, we derive exact relations between
the QCD$_3$ partition function and the  QCD$_4$ partition function
continued to half-integer topological charge.
The replica limit of these formulas
results in exact relations between the corresponding
microscopic spectral densities of QCD$_3$ and QCD$_4$. Replica
calculations, which are exact for QCD$_4$ at half-integer topological charge,
thus result in exact expressions for the  microscopic
spectral density of the QCD$_3$ Dirac operator.
Second, we derive Virasoro constraints for
the QCD$_3$ partition function. They uniquely
determine the small-mass expansion of the partition function
and the corresponding sum rules for inverse Dirac eigenvalues.
Due to de Wit-'t Hooft poles,
the replica limit only reproduces the small mass expansion
of the resolvent up to a finite number of terms.
Third, the large mass expansion of the resolvent is obtained from the replica
limit of a loop expansion of the QCD$_3$ partition function.
Because of Duistermaat-Heckman localization
exact results are obtained for the
microscopic spectral density in this way.
\end{abstract}
\vfill

\begin{flushleft}
NBI-HE-00-38 \\
SUNY-NTG-00/54\\
hep-th/0011072
\end{flushleft}
\thispagestyle{empty}
\newpage

\setcounter{equation}{0}
\section{Introduction}
\vspace*{0.3cm}

\noi
QCD in 2+1 space-time dimensions (QCD$_3$) provides an
interesting alternative to spontaneous chiral symmetry breaking
in 3+1 dimensions. Because of the absence of a $\gamma_5$ matrix
in any odd number of
space-time dimensions, ordinary chiral symmetry cannot be defined.
Yet, for an even number of flavors $2N_f$ it has been known for
some time that there exists a close analogue of chiral symmetry in
2+1 dimensions (see, $e.g.$, ref. \cite{P}).
In this case the  global flavor symmetry may break spontaneously according to
$U(2N_f) \to U(N_f)\times U(N_f)$ \cite{P,VZ}. For an odd number
of flavors it has been argued that the pattern of spontaneous
symmetry breaking pattern is replaced by $U(2N_f+1) \to U(N_f+1)\times
U(N_f)$ \cite{VZ}. This picture of spontaneous flavor symmetry
breaking  has been supported by recent Monte Carlo simulations \cite{DHKM}.

\noi
Spontaneous flavor symmetry breaking such as discussed above gives rise to
one Goldstone boson for each broken generator.
In ref. \cite{VZ}
the low-energy effective partition function of the corresponding
chiral Lagrangian was used to extract
information on the spectrum of the Dirac operator of QCD$_3$.
It was also suggested \cite{VZ} that Random Matrix Theory may be
used to exactly compute the microscopic properties of the Dirac
spectrum in much the same way as was first proposed
for QCD in 3+1 dimensions (QCD$_4$) \cite{SV}.
The resulting spectral correlation functions for the ensemble relevant
for QCD$_3$ (a variant of the Unitary Ensemble, UE) are indeed
universal in Random Matrix Theory \cite{ADMN,DN,Jesper}.
They can be expressed entirely
in terms of (2+1)-dimensional finite-volume partition
functions \cite{AD1}.
Recently, the patterns of flavor symmetry breaking for theories with
gauge group SU(2) or with fermions in the adjoint representation of
SU($N_c$) in 2+1 dimensions have been shown to correspond
to two other classes of the classical Random Matrix Theories \cite{M},
namely the Orthogonal Ensemble (OE) and the Symplectic Ensemble (SE),
respectively, in precise analogy to the (3+1)-dimensional case \cite{JJMV}.
The microscopic spectral densities
of these theories were calculated recently in \cite{HN,NN}.

\noi
To derive non-perturbative analytical results for the smallest eigenvalues
of the Dirac operator it is important to {\em prove} that universal
Random Matrix Theory results coincide with
the low-energy limit of the field theory. For QCD$_4$
this has been achieved for the spectral density and the two-point spectral
correlation function by means of the
so-called supersymmetric method
\cite{OTV,TV2,Brezin,Efetov,VWZ,simons-altland,Takahashi,Guhr-Wilke-Weidi}.
In this method, one
adds  a set of fermionic and bosonic ghost quarks (or ``valence quarks'',
in a terminology borrowed from lattice gauge theory
\cite{Maarten,Leung,Sharpe})
such that their determinants cancel for equal masses. In this way
one obtains a generating functions for
the resolvent (the partially quenched chiral condensate
in the language of lattice gauge theory) or the higher
order correlations functions. The spectral density then follows
from the discontinuity of the resolvent in the complex mass plane of the
ghost quarks \cite{OTV}.

\noi
An alternative method to eliminate the
determinants of ghost quarks is the replica method \cite{EA}. It is based on
adding  either $n$ fermionic or $n$ bosonic valence quarks
(also called
replicas),
performing an analytical continuation in $n$                            
and taking the limit $n \to 0$ at the end of the calculation.
Although the replica method has been challenged                      
\cite{Vzirn},
this method has received  a considerable boost in the context of
condensed matter physics during the past year
\cite{Mezard,lerner,kanzieper,Z1}.
Recently, the
replica method has also been applied to the field theory corresponding
to the low-energy limit of QCD$_4$
\cite{DS,DV,Berb99}. The advantage of the fermionic replica method is that we
are familiar with the pattern of spontaneous chiral symmetry breaking, and
it is straightforward to define the corresponding low-energy field theory.
Although we have less experience with spontaneous symmetry breaking in
the supersymmetric formulation, the consensus is that such internal
supersymmetries are not spontaneously broken. Based on this assumption
there should not be any problems in defining the low-energy limit of the
supersymmetric partition function of this theory, which in turn
is the generating function of the resolvent. This has recently
been confirmed by the explicit calculation of Szabo \cite{Sz}, who has
demonstrated how the microscopic spectral density and other spectral
correlation functions of QCD$_3$ can be obtained in this way.
Nevertheless, precisely in the
case of QCD$_3$ it is advantageous to explore also the replica
method. The reason is as follows. In general, one expects that the replica
method can only provide us
with asymptotic series for spectral correlation functions of the Dirac
operator, and not with exact results. In QCD$_3$, however,
the large mass asymptotic expansion of the
resolvent terminates (it
is semiclassically exact \cite{Z1} in the quenched case)
and exact results will be obtained this way.

\noi
An important ingredient of this paper is the discussion of
general relations between the
finite-volume partition functions for QCD in 2+1 and 3+1 dimensions. They
will be introduced in section 2 and the relations will be formulated and
proved in section 3.
These relations
hold in all generality and are
independent of any detailed derivations in any of the two theories.
The implications of these relations are discussed in section 4, where
we use the replica method to derive our main result: two completely general
relations between the microscopic spectral densities of the two theories.
Thus, once we know the spectral densities of QCD$_4$, the spectral
densities of QCD$_3$ follow. In section 4.1 it is shown that results for
the microscopic spectral density and microscopic
spectral correlation functions obtained in this way
are in exact agreement with previous results
form Random Matrix Theory \cite{VZ}.
An exact calculation
of the resolvent of the QCD$_3$ partition function from its finite volume
partition function is thus possible by using the recent results for
the QCD$_4$ finite volume partition function
\cite{DV} (see section 4.2).
In sections 5 and 6  we perform explicit
calculations in QCD$_3$ based on the replica method. First, in section 5,
a small-mass
series expansion of the resolvent of the QCD$_3$ Dirac operator is
generated by means of Virasoro constraints on the
finite volume QCD$_3$ partition function.
As a by-product of this analysis we derive a series of
spectral sum rules for  the Dirac operator in QCD$_3$. In section 6, we
explore the (large-mass) saddle-point expansion of the
resolvent,
which turns out to {\em truncate} after a finite number of terms,
and hence is exact. This truncation is consistent with the general
relations to the QCD$_4$ partition function
derived in section 3.
Our results are summarized in section 7.

\setcounter{equation}{0}
\section{The QCD$_3$ Partition Function}
\vspace*{0.3cm}

\noi
The Dirac operator of the QCD$_3$ partition function
is given by
\be
D= i(\partial_k + iA_k) \sigma_k ,
\ee
where the $\sigma_k$ are the Pauli matrices and the $A_k$ are the
$SU(N_c)$ gauge fields. The partition function for $2N_f$ flavors
is defined by
\be
{\cal Z} = \langle \det (D {\bf 1}_{2N_f} + iM ) \rangle,
\ee
where $M$ is a Hermitian mass matrix and the average is over
the gauge field configurations weighted by the Yang-Mills action.
Without loss of generality, the mass matrix can always be chosen diagonal, and
we will only consider this case.
Generally, the fermion determinant in QCD$_3$ is not positive definite,
resulting in a Chern-Simons term due to the anomaly \cite{Red}.
However, for quark
masses occuring in pairs $m_k$ and $-m_k$, the fermion determinant is
positive definite, and the theory is anomaly free. In that case the
sign of the $\langle \bar \psi_k \psi_k \rangle$ is determined by
the sign of the quark mass,
\be
\langle \bar \psi_k \psi_k \rangle = -\frac {m_k}{|m_k|} \Sigma.
\label{cond3}
\ee

\noi
The low-energy effective partition function is constructed from the
requirements that its transformation properties should reflect those
of QCD$_3$, i.e. the partition function is invariant under $U\in U(2N_f)$
transformations of the quark fields if, at the same time, the mass matrix
is transformed according to
\be
M \to U M U^{-1}.
\ee
The chiral condensate given by (\ref{cond3}) is only invariant under
an $U(N_f) \times U(N_f)$ subgroup of $U(2N_f)$. The Goldstone manifold
is thus given by $U(2N_f)/U(N_f) \times U(N_f)$.  The matrices
in this manifold transform according to
\be
U \to U_1 U U_1^{-1}, \qquad U_1 \in U(2N_f).                               
\label{vunit}
\ee
To lowest order in the mass, the mass term of the effective theory is
given by
\be
{\cal L}_m = \pm {\rm Re}
\Sigma {\rm Tr} M U.
\ee
Because of the integration over the coset in the partition function,
the overall sign of the  mass term is irrelevant.
For an even number of flavors
in the limit where the Compton wavelength of the pseudo-Goldstone bosons 
is much larger than the size of the box,                                 
the partition
function factorizes into a zero momentum part and a non-zero
momentum part. The first one, which is also known as the QCD$_3$
finite volume partition
function, is thus given by \cite{VZ}
\beq
{\cal Z}_{{\rm QCD}_3}^{(2N_f)}(\{\mu_i\})\ =\ \int_{U(2N_{f})}\! dU \exp
[V\Sigma {\mbox{\rm Tr}({\cal M}}U\Gamma_5 U^\dagger)]\ ,
\label{ZQCD3even}
\eeq
with  ${\cal M}$=diag$(m_1,\ldots,m_{N_f},-m_1,\ldots,-m_{N_f})$\footnote{
Recall that in 2+1 dimensions there is no definite sign for the fermionic
mass term, a consequence of the fact that in any odd number of space-time
dimensions the two sets of $\gamma$-matrices $\{\gamma_i\}$ and
$\{-\gamma_i\}$ are inequivalent irreducible representations of the
Clifford algebra.},
$\Gamma_5$=diag({\bf 1}$_{N_f}$,-{\bf 1}$_{N_f}$) and $V$ the volume   
of the three-dimensional space-time.                                   
We have used an explicit representation of the coset, $U \to U\Gamma_5
U^\dagger$, and extended the integration to the full $U(2N_f)$ group. This
gives only rise to an overall volume factor.
The partition function (\ref{ZQCD3even}) is only a function of the
rescaled masses
\be
\mu_i \equiv m_i V \Sigma,
\ee
which will be kept fixed as $V \to \infty$.
We shall here consider QCD$_3$
at fixed ultraviolet cut-off $\Lambda$. In the infrared the theory is
regularized by the three-volume $V$, which is only eventually taken to
infinity.
If not only the volume is finite but also
space-time is discretized such as in lattice QCD, the total number
of Dirac eigenvalues is finite and will be denoted by ${\cal N}$.

\noi
For an odd number of flavors one of the quarks is unpaired, and the fermion
determinant may change sign under large gauge transformations. 
In this case the sign of the
chiral condensate is not given by the sign of the quark mass. There
are two inequivalent possibilities for the vacuum state,
determined by the sign of the fermion
determinant. They are not connected by unitary transformations of the
coset elements.
In a regime for which $m \sim 1/V
\Sigma$, both states have to be included in the
low energy limit of the QCD partition function. The QCD$_3$ partition  function
satisfies the parity relation
\be
{\cal Z}_{{\rm QCD}_3}(-m) = (-1)^{{\cal N}N_f}{\cal Z}_{{\rm QCD}_3}(m).
\ee
For an odd number of flavors the low-energy effective partition function
is thus given by
\be
{\cal Z}_{{\rm QCD}_3^+}^{(2N_f+1)}(\{\mu_i\})\ =\
\int_{U(2N_{f}+1)}\! dU
\cosh[V\Sigma {\mbox{\rm Tr}({\cal M}}U \tilde\Gamma_5 U^\dagger)]\ ,
\label{ZQCD3odd0}
\ee
for even number of eigenvalues ${\cal N}$, and
\be
{\cal Z}_{{\rm QCD}_3^-}^{(2N_f+1)}(\{\mu_i\})\ =\
\int_{U(2N_{f}+1)}\! dU
\sinh[V\Sigma {\mbox{\rm Tr}({\cal M}}U\tilde \Gamma_5 U^\dagger)]\ ,
\label{ZQCD3odd1}
\ee
for odd number of eigenvalues.
The second partition function (\ref{ZQCD3odd1}) is not positive definite,
and is unsuitable as a physical partition function. In particular, it
vanishes when the unpaired fermion mass vanishes \cite{Jesper}.
Nevertheless, this second partition function can appear at intermediate
stages \cite{AD1}.

\noi
These partition functions are to be compared with the QCD$_4$ finite volume
partition function given by \cite{SV}
\beq
{\cal Z}^{(N_f)}_{\nu}(\{\mu_i\}) ~=~ \int_{U(N_f)}\! dU~ (\det U)^{\nu}
\exp\left[\frac{1}{2}V\Sigma {\rm Tr}({\cal M}
U^{\dagger} + U{\cal M}^{\dagger})\right] ~, \label{ZchUE}
\eeq
where $\nu$ is the topological charge, and
${\cal M}$=diag$(m_1,\ldots,m_{N_f})$. In both cases
$\Sigma$ stands for the infinite-volume quark-antiquark condensate.
As we will see in the next section,  there are deep relations between
these two types of group integrals.
Although the structural nature of these relations suggest that they  
can be obtained from general group theoretical arguments, we have only  
been able to prove them at the technical level.                         

\setcounter{equation}{0}
\section{The QCD$_3$--QCD$_4$ Connection}
\vspace*{0.3cm}

\noi
In this section we shall prove a set of surprising relations between
group integrals of Itzykson-Zuber kind (\ref{ZQCD3even}),
(\ref{ZQCD3odd0}) and  (\ref{ZQCD3odd1}),
and group integrals  of ``external field'' type (\ref{ZchUE}).
Some of these relations were already pointed out in ref. \cite{AD2},
based on the relationship to Random Matrix Theory, but we shall
here prove them directly from the finite volume partition functions.

\noi
Both of the group integrals in eqs. (\ref{ZQCD3even}) and (\ref{ZchUE})
can be expressed in closed form. The first is simply an Itzykson-Zuber
integral \cite{IZ} with a particular $N_f$-degeneracy in one of the
matrices ($\Gamma_5$), and the result is \cite{DN} \footnote{Note
  that we have chosen a different prefactor than in ref. \cite{DN} in
  order to make the partition function positive definite. For averages
  the difference is of course irrelevant.}
\beq
{\cal Z}^{(2N_{f})}_{{\rm QCD}_3}(\{\mu_i\}) ~=~ (-1)^{N_f(N_f+1)/2}
\ \frac{\det\left(
\begin{array}{ll}
\matA(\{\mu_i\}) & \matA(\{-\mu_i\})\\
\matA(\{-\mu_i\})  & \matA(\{\mu_i\})
\end{array}\right)}{\Delta({\cal M})} ~,\label{ZQCD3evenexpl}
\eeq
where the $N_f\!\times\!N_f$ matrix
$\matA(\{\mu_i\})$ is defined by
\beq
\matA(\{\mu_i\})_{jl} ~\equiv~ (\mu_j)^{l-1} \mbox{e}^{\mu_j}
~~,~~~~j,l=1,\ldots,N_f~.
\label{matAdef}
\eeq
and the Vandermonde determinant is taken over all masses (including
the sign-mirrored ones),
\beq
\Delta({\cal M}) ~=~ \prod_{i>j}^{2N_{f}}(\mu_i-\mu_j) ~.
\eeq
Recently, it was shown that the above partition function can be
expressed much more compactly by means of just a single $N_f\times
N_f$ determinant \cite{Sz}. However, precisely for our present
purposes the above form is actually more convenient.

\noi
For an odd number of flavors the integral (\ref{ZQCD3odd0}) and
(\ref{ZQCD3odd1}) can also be evaluated as Itzykson-Zuber integrals. The
result is
\beq
{\cal Z}^{(2N_{f}+1)}_{{\rm QCD}_3^\pm}(\mu,\{\mu_i\}) ~=~ (-1)^{N_f(N_f+3)/2}
\frac{2^{N_f}}{\Delta({\cal M})}\frac12
\left[ \det \matD(\mu,\{\mu_i\}) \pm (-1)^{N_f}
\det \matD(-\mu,\{-\mu_i\})\right]~,
\label{ZQCD3oddexpl}
\eeq
where the $(2N_f+1)\times(2N_f+1)$ matrix $\matD$ is defined as
\beq
2^{N_f} \det \matD(\mu,\{\mu_i\}) ~\equiv~
\det\left(
\begin{array}{ll}
\matA(\mu,\{\mu_i\})_{N_f+1\times N_f+1}
& \matA(-\mu,\{-\mu_i\})_{N_f+1\times N_f}\\
\matA(\{-\mu_i\})_{N_f\times N_f+1}
& \matA(\{\mu_i\})_{N_f\times N_f}
\end{array}\right).
\label{matDdefodd}
\eeq
Here, the matrix $\matA$ is given by eq. (\ref{matAdef}), with the
corresponding quadratic or rectangular size explicitly indicated.

\noi
For integer $\nu$ the  integral (\ref{ZchUE})                     
can be expressed as \cite{JSV},                                   
\beq
{\cal Z}_\nu^{(N_f)} (\{\mu_i\}) ~=~ \det \matB(\{\mu_i\})/\Delta(\{\mu_i^2\}),
\label{ZchUEexpl}
\eeq
where the matrix $\matB$ in eq. (\ref{ZchUEexpl}) is given by
\beq
\matB(\{\mu_i\})_{jl}
= \mu_j^{l-1}I_{\nu}^{(l-1)}(\mu_j)~~,~~~~j,l=1,\ldots,N_f ~,
\label{matBdef}
\eeq
with $I_{\nu}^{(l-1)}(\mu )= (\partial /\partial \mu )^{l-1}I_{\nu}(\mu )$. 
The denominator is given by the Vandermonde determinant of
squared masses,
\beq
\Delta (\{\mu_i^2\})\ \equiv\ \prod_{i>j}^{N_f}(\mu_i^2-\mu_j^2)
\ =\ \det_{i,j}\left[ (\mu_i^2)^{j-1}\right] \ .
\label{Vandermonde}
\eeq
In both cases the normalization convention of the
integrals has been chosen for later convenience.

\noi
For non-integer values of $\nu$ we define the QCD$_4$ partition function
according to the analytical continuation in $\nu$ of the Bessel
functions \cite{DV},
\be
I_\nu(z) =\frac 1{2\pi} \int_{-\pi}^{\pi}
e^{z\cos \theta} e^{i\nu \theta} d\theta
-\frac{\sin \nu \pi}\pi \int_0^\infty ds e^{-z\cosh s -\nu s},
\label{besselnu}
\ee
and not by analytical continuation in $\nu$ of
the unitary matrix integral (\ref{ZchUE}).

The integral (\ref{ZchUE}) satisfies an important relation known 
as flavor-topology duality \cite{Jac}. A proof that greatly   
simplifies earlier proofs \cite{Jac}                             
is obtained by                                                 
realizing that the matrix ${\bf B }$ in (\ref{ZchUEexpl})  can 
be replaced by                                                  
\beq                                                   
\matB(\{\mu_i\})_{jl}                                  
\to \mu_j^{l-1}I_{\nu+l-1}(\mu_j)~~,~~~~j,l=1,\ldots,N_f ~, 
\label{matBdef2}                                             
\eeq                                                         
If we take the limit $\mu\to 0$ and use the fact that         
$I_{\nu}(\mu)\sim (\mu/2)^\nu / \Gamma(\nu+1)$ we find the     
the flavor-topology relation                                  
 \beq                                                         
\lim_{\mu\to 0} \Gamma(\nu+1)\left(\frac{2}{\mu}\right)^\nu       
{\cal Z}_\nu^{(N_f+1)}(\mu,\{\mu_i\})                             
~=~                                                                   
\prod_{j=1}^{N_f}\frac{1}{\mu_j}\ {\cal Z}_{\nu+1}^{(N_f)}(\{\mu_i\}) 
~.                                       
\label{FTfactornu}                    
\eeq                                 
The general relation  obtained by iterating (\ref{FTfactornu}) 
with repect to $N_f$ 
is valid for arbitrary $\nu$ provided that ${\cal Z}_\nu^{(N_f)}$ 
is analytically continued in $\nu$ according to (\ref{besselnu}).  

\noi
After these preliminary definitions, we are now ready to state and prove
three theorems.
\vspace{0.5cm}

\newpage
\noi\noindent
{\sc Theorem I} - Even number of flavors in QCD$_3$:

\noi
{\it Let ${\cal Z}_{{\rm QCD}_3}^{(2N_f)}(\{\mu_i\})$ and
${\cal Z}^{(N_f)}_{\nu}(\{\mu_i\})$ be as defined in eqs. (\ref{ZQCD3even})
and (\ref{ZchUE}), with the normalization conventions
(\ref{ZQCD3evenexpl}) -- (\ref{Vandermonde}). Then the following
identity holds:
\beq
{\cal Z}_{{\rm QCD}_3}^{(2N_f)}(\{\mu_i\}) ~=~
\pi^{N_f}{\cal Z}_{\nu=-1/2}^{(N_f)}(\{\mu_i\})
                   ~ {\cal Z}_{\nu=+1/2}^{(N_f)}(\{\mu_i\}) ~.
\label{factoreven}
\eeq }

\noi\noindent
{\sc Proof}:
We start by simplifying the right hand side, using that for
$\nu=\pm 1/2$ we have
\beq
I_{\nu=-1/2}(\mu) ~=~ \sqrt{\frac{2}{\pi\mu}}\cosh(\mu)
~~~~ \mbox{and} ~~~~
I_{\nu=+1/2}(\mu) ~=~ \sqrt{\frac{2}{\pi\mu}}\sinh(\mu)
~.\label{halfI}
\eeq
One can easily convince oneself that in
eq. (\ref{matBdef})  only terms with all derivatives acting on $\cosh(\mu_j)$
and $\sinh(\mu_j)$, respectively, contribute to the partition
function.
All other terms containing
derivatives of the square roots can be
eliminated by adding multiples of columns from the right, which leaves
the determinant invariant. We can thus pull out the common factors
of square roots and arrive at
\beqn
{\cal Z}_{\nu=-1/2}^{(N_f)}(\{\mu_i\}) &=&
\prod_{j=1}^{N_f}\sqrt{\frac{2}{\pi\mu_j}}
\det \matC(\{\mu_i\})/ \Delta(\{\mu_i^2\}) \ ,\ \ \
\matC(\{\mu_i\})_{jl}\equiv\mu_j^{l-1}\cosh^{(l-1)} (\mu_j)
\nonumber\\
{\cal Z}_{\nu=+1/2}^{(N_f)}(\{\mu_i\}) &=&
\prod_{j=1}^{N_f}\sqrt{\frac{2}{\pi\mu_j}}
\det \matS(\{\mu_i\})/ \Delta(\{\mu_i^2\})  \ ,\ \ \
\matS(\{\mu_i\})_{jl}\equiv
\mu_j^{l-1}\sinh^{(l-1)} (\mu_j)
~.\label{rhseven}
\eeqn
Here and below $\cosh^{(p)} (\mu)$ or $\sinh^{(p)}(\mu)$ denotes the 
$p$-th derivative of $\cosh(mu)$ or $\sinh(mu)$.                     
Turning next to the QCD$_3$ side, the Vandermonde determinant can    
be written as,
\beq
\Delta({\cal M}) ~=~
(-1)^{N_f(N_f+1)/2}
\prod_{k=1}^{N_f} 2\mu_k \ \Delta^2(\{\mu_i^2\})
~.\label{Deltarel}
\eeq
In order to bring the $(2N_f)\times(2N_f)$ determinant in eq.
(\ref{ZQCD3evenexpl}) into the desired form we iteratively add and
subtract columns to obtain $\cosh$ and $\sinh$.
Starting with the first column we
add the $N_f$-th column to obtain
$(2\cosh(\mu_1),\ldots,
2\cosh(\mu_{N_f}),2\cosh(\mu_1),\ldots,2\cosh(\mu_{N_f}))$.
Then, subtracting 1/2 of this new first column from the $N_f$-th
column, we obtain $(-\sinh(\mu_1),\ldots,
-\sinh(\mu_{N_f}),\sinh(\mu_1),\ldots,\sinh(\mu_{N_f}))$ as a new
column vector at position $N_f$. The common
factor of 2 can be taken out of the first column. We proceed in the
same way with
the second and $(N_f+2)$-nd column, where $\cosh$ and $\sinh$ get
interchanged in the resulting columns compared to the previous step.
After completing this alternating procedure we arrive at the
following block structure for the QCD$_3$ side:
\beqn
{\cal Z}_{{\rm QCD}_3}^{(2N_f)}(\{\mu_i\}) &=& (-1)^{N_f(N_f+1)/2}
\frac{2^{N_f}}{\Delta({\cal M})}
\det\left(
\begin{array}{ll}
\matC(\{\mu_i\}) & -\matS(\{\mu_i\})\\
\matC(\{\mu_i\})  & \ \ \matS(\{\mu_i\})
\end{array}\!\right) \nn\\
&=&
(-1)^{N_f(N_f+1)/2}
\frac{2^{N_f}}{\Delta({\cal M})}
\det\left(
\begin{array}{ll}
2\matC(\{\mu_i\}) & \ \ \mbox{\bf 0}\\
\ \matC(\{\mu_i\})  & \matS(\{\mu_i\})
\end{array}\right)
~,\label{ZQDC3block}
\eeqn
where in the last step we have added the lower blocks to the
upper ones, by a subsequent addition of rows. We can now perform a
Laplace expansion into products of $N_f\times N_f$ blocks,
which due to the
the zero block, {\bf 0}, only yields one single
product of two $N_f\times N_f$
determinants. Together with eq. (\ref{Deltarel}) we obtain the final result:
\beq
{\cal Z}_{{\rm QCD}_3}^{(2N_f)}(\{\mu_i\}) ~=~
\frac{2^{N_f}}{\Delta^2(\{\mu_i^2\})
\prod_{k=1}^{N_f}\mu_k}
\det \matC(\{\mu_i\}) \det\matS(\{\mu_i\})
~.\label{ZQDC3final}
\eeq
By comparing eqs. (\ref{ZQDC3final}) and (\ref{rhseven}) we can read
off the prefactor in the theorem eq. (\ref{factoreven}).

\vspace{0.5cm}
\noi\noindent
{\sc Theorem II} - Odd number of flavors in QCD$_3$:

\noi
{\it Let in addition ${\cal Z}_{{\rm QCD}_3^\pm}^{(2N_f+1)}(\mu,\{\mu_i\})$
be as defined in eqs. (\ref{ZQCD3oddexpl}) and (\ref{matDdefodd}).
Then the following two identities hold:
\beq
{\cal Z}_{{\rm QCD}_3^\pm}^{(2N_f+1)}(\mu,\{\mu_i\}) ~=~
\pi^{N_f}
\sqrt{\frac{\pi\mu}{2}}\ {\cal Z}_{\nu=\mp 1/2}^{(N_f+1)}(\mu,\{\mu_i\})
            \ {\cal Z}_{\nu=\pm 1/2}^{(N_f)}(\{\mu_i\}) \ .
\label{factorodd}
\eeq}

\noi\noindent
{\sc Proof}: Since the right hand side already follows
from eq. (\ref{rhseven}) we start with the QCD$_3$ side, treating both
cases, $\pm$, simultaneously. The Vandermonde determinant
with one additional mass
can be treated similarly to relation (\ref{Deltarel}) and we obtain
\beq
\Delta({\cal M}) ~=~
(-1)^{N_f(N_f+1)/2}
\prod_{j=1}^{N_f} (\mu^2-\mu_j^2)
\prod_{l=1}^{N_f} 2\mu_l\ \Delta(\{\mu_i^2\})^2
~.\label{Deltarelodd}
\eeq
Next, we  proceed by adding and subtracting columns as
in the proof of the previous theorem. We arrive at the same
determinant with blocks of matrices $\matC$ and $\matS$, except that
the $(N_f+1)$-th columns remains unchanged. Here, we are dealing with an
odd number of flavors and the determinant in eq. (\ref{ZQCD3oddexpl})
is of size $(2N_f+1)\times (2N_f+1)$, with leaves one column unpaired.
We thus have for eq. (\ref{matDdefodd})
\beq
2^{N_f} \det \matD(\mu,\{\mu_i\}) ~=~ 2^{N_f}
\det\left(
\begin{array}{lll}
\mu^{j-1}\cosh^{(j-1)} (\mu)&\ \ \mu^{N_f}\mbox{e}^{\mu}
&-\mu^{j-1}\sinh^{(j-1)} (\mu) \\
\matC(\{\mu_i\})  & \ \ \mu_i^{N_f}\mbox{e}^{\mu_i} & -\matS(\{\mu_i\})\\
\matC(\{\mu_i\})  & (-\mu_i)^{N_f}\mbox{e}^{-\mu_i} &\ \ \matS(\{\mu_i\})
\end{array}\right).
\eeq
When adding the second determinant of eq. (\ref{ZQCD3oddexpl}) with
negative arguments we can use the fact that the matrix $\matC$ is even
whereas  $\matS$ is odd. After taking out all the minus signs from the
matrix $\matS$, which gives $(-1)^{N_f}$, we can add the two determinants
$\det \matD(\mu,\{\mu_i\})$ and $\det \matD(-\mu,\{-\mu_i\})$, as they
now differ only by the $(N_f+1)$-th column.
The resulting determinant reads
\beqn
\det \matD(\mu,\{\mu_i\}) &\pm& (-1)^{N_f}\det
\matD(-\mu,\{-\mu_i\})~=
\label{Dinter}\\
&=&
\det\left(
\begin{array}{lll}
\mu^{j-1}\cosh^{(j-1)} (\mu)&
\ \ \mu^{N_f}(\mbox{e}^{\mu}\pm(-1)^{N_f} \mbox{e}^{-\mu})
&-\mu^{j-1}\sinh^{(j-1)} (\mu) \\
\matC(\{\mu_i\}) &
\ \ \mu_i^{N_f}(\mbox{e}^{\mu_i}\pm(-1)^{N_f} \mbox{e}^{-\mu_i})
& -\matS(\{\mu_i\})\\
\matC(\{\mu_i\})  &
\pm \mu_i^{N_f}(\mbox{e}^{\mu_i}\pm(-1)^{N_f} \mbox{e}^{-\mu_i})
&\ \ \matS(\{\mu_i\})\end{array}\right). \nn
\eeqn
We now have to treat each of the two cases $\pm$ separately.
For the ``+'' case the $(N_f+1)$-th column reads
$(2\mu^{N_f}\!\cosh^{(N_f)}(\mu),2\mu_1^{N_f}\!\cosh^{(N_f)} (\mu_1),\ldots,
2\mu_1^{N_f}\!\cosh^{(N_f)} (\mu_1),\ldots)$
which can be absorbed into the matrices $\matC$.
The ``--'' case yields
$(2\mu^{N_f}\!\sinh^{(N_f)}(\mu),2\mu_1^{N_f}\!\sinh^{(N_f)}(\mu_1),\ldots,
-2\mu_1^{N_f}\!\sinh^{(N_f)} (\mu_1),
\ldots)$,
and we can absorb it into the
matrices $\matS$ after changing the sign of the vector and
permuting it to the very right.
We will then add or subtract blocks again to produce a
{\bf 0}$_{N_f\times N_f}$ block
on the right or left, respectively. In more detail we have for the
``+'' case
\beqn
\det \matD(\mu,\{\mu_i\})+(-1)^{N_f}\det \matD(-\mu,\{-\mu_i\})&=&\\
&=& 2
\det\left(
\begin{array}{ll}
\mu^{j-1}\cosh^{(j-1)}(\mu) &-\mu^{j-1}\sinh^{(j-1)} (\mu)\\
\matC(\{\mu_i\})_{N_f\times N_f+1} & -\matS(\{\mu_i\})_{N_f\times N_f}\\
\matC(\{\mu_i\})_{N_f\times N_f+1}  &\ \ \matS(\{\mu_i\})_{N_f\times  N_f}
\end{array}\!\right) \nn\\
&=& 2
\det\left(
\begin{array}{ll}
\mu^{j-1}\cosh^{(j-1)} (\mu) &-\mu^{j-1}\sinh^{(j-1)}(\mu) \\
2\matC(\{\mu_i\})_{N_f\times N_f+1} & \mbox{\bf 0}_{N_f\times N_f}\\
\ \matC(\{\mu_i\})_{N_f\times N_f+1}  &\matS(\{\mu_i\})_{N_f\times N_f}
\end{array}\!\!\right). \nn
\eeqn
Because of the zero block {\bf 0}$_{N_f\times N_f}$,
a Laplace expansion into products of $(N_f+1)\times(N_f+1)$
blocks and $N_f\times N_f$ blocks contains at most $N_f+1$ nonzero terms.
However, all the determinants for which two
rows of the matrices $\matC$ occur twice, also vanish. Thus we are left with
\beq
\det\matD(\mu,\{\mu_i\})+(-1)^{N_f}\det \matD(-\mu,\{-\mu_i\})~=~
2^{N_f+1}\det\matC(\mu,\{\mu_i\}) \det \matS(\{\mu_i\})\ ,
\label{Deven}
\eeq
where the size of the determinants on the right hand side
is given by the number of arguments.

Along the same lines we obtain for the ``--''-case,
\beqn
\det \matD(\mu,\{\mu_i\}) -
(-1)^{N_f}\det \matD(-\mu,\{-\mu_i\}) &&=
\label{Dodd}\\
=&&\!\!\!\!(-1)^{N_f+1}
\det\!\left(\!
\begin{array}{ll}
\mu^{j-1}\cosh^{(j-1)} (\mu) &-\mu^{j-1}\sinh^{(j-1)} (\mu) \\
\matC(\{\mu_i\})_{N_f\times N_f} & -\matS(\{\mu_i\})_{N_f\times N_f+1}\\
\matC(\{\mu_i\})_{N_f\times N_f}  &\ \ \matS(\{\mu_i\})_{N_f\times  N_f+1}
\end{array}\!\right) \nn\\
=&&\!\!\!\!(-1)^{N_f+1}
\det\!\left(\!
\begin{array}{ll}
\mu^{j-1}\cosh^{(j-1)}(\mu) &-\mu^{j-1}\sinh^{(j-1)}(\mu) \\
\mbox{\bf 0}_{N_f\times N_f}& -2\matS(\{\mu_i\})_{N_f\times N_f+1} \\
\matC(\{\mu_i\})_{N_f\times N_f} &\ \ \ \ \matS(\{\mu_i\})_{N_f\times N_f+1}
\end{array}\!\!\!\right) \nn\\
=&&
\!\!\!\!2^{N_f+1}\det\matC(\{\mu_i\}) \det \matS(\mu,\{\mu_i\})\ .\nn
\eeqn
In the last step the minus signs coming from permuting the
$(N_f+1)$-th column of eq. (\ref{Dinter}) get absorbed into the sign of
the only non-vanishing contribution from the Laplace expansion. Taking
eqs. (\ref{Deven}) and (\ref{Dodd}) together with
eq. (\ref{Deltarelodd}) we finally obtain
\beq
{\cal Z}_{{\rm QCD}_3^\pm}^{(2N_f+1)}(\mu,\{\mu_i\})~=~
\frac{(-1)^{N_f}\ 2^{N_f} }{\Delta(\{\mu_i^2\})^2
\prod_{k=1}^{N_f}(\mu^2-\mu_k^2)\mu_k}
\left\{ \begin{array}{ll}
\det\matC(\mu,\{\mu_i\})\det \matS(\{\mu_i\}) & \mbox{for} \ + \\
\det\matS(\mu,\{\mu_i\})\det \matC(\{\mu_i\}) & \mbox{for} \ -
\end{array}\right.~.
\eeq
We can now compare to the right hand side of eq. (\ref{factorodd})
with the help of eq. (\ref{rhseven}). Looking at the factor $\prod_k
(\mu_k^2-\mu^2)$ of the Vandermonde determinant
of ${\cal Z}^{(N_f+1)}_\nu(\mu,\{\mu_i\})$
we obtain an additional factor $(-1)^{N_f}$, which adds up to the
correct prefactor of eq. (\ref{factorodd}).

\noi
The first two theorems stated so far relate the QCD$_3$
partition function with an odd or even number of flavors to a single
product of two QCD$_4$ partition functions. In the following we wish
to consider the QCD$_3$ partition function with additional flavors
which do not come in pairs of opposite sign. However, such partition
functions do not satisfy relations as simple as Theorem I and II.
Instead, they are given
by linear combinations of products of QCD$_4$ partition
functions and do not directly give relations between the spectral
density of QCD$_3$ and QCD$_4$.
For this reason we will only treat the simplest case,
that of the even flavor QCD$_3$ partition function with two additional
unpaired masses.
\vspace{0.5cm}

\noi\noindent
{\sc Theorem III} - Additional unpaired flavors in QCD$_3$:

\noi
{\it Let in addition
${\cal Z}^{(2N_{f}+2)}_{{\rm QCD}_3}(\{\mu_i\},\zeta,\om)$ be defined as}
\beq
{\cal Z}^{(2N_{f}+2)}_{{\rm QCD}_3}(\{\mu_i\},\zeta,\om) ~=~
(-1)^{(N_f+1)(N_f+2)/2}
\ \frac{\det\left(
\begin{array}{ll}
\matA(\{\mu_i\},\zeta) & \matA(\{-\mu_i\},-\zeta)\\
\matA(\{-\mu_i\},\om)  & \matA(\{\mu_i\},-\om)
\end{array}\right)}{\Delta({\cal M})} ~,\label{ZQCD3zw}
\eeq
{\it where the $(N_f+1)\!\times\!(N_f+1)$ matrix
$\matA(\{\mu_i\},\zeta)$ is defined as in eq. (\ref{matAdef}).
The mass matrix is given by
${\cal M}\!=\!
\mbox{\rm diag}(\mu_1,\ldots,\mu_{N_f},\zeta,-\mu_1,\ldots,-\mu_{N_f},\om)$.
Then the two following relations hold:}
\beqn
{\cal Z}^{(2N_{f}+2)}_{{\rm QCD}_3}(\{\mu_i\},\pm\zeta,\om) &=&
\pi^{N_f+1} \sqrt{\zeta\om}
\left[
{\cal Z}^{(N_{f}+2)}_{\nu=-1/2}(\{\mu_i\},\zeta,\om) \
{\cal Z}^{(N_{f})}_{\nu=+1/2}(\{\mu_i\}) \right.\nn \\
&&\ \ \ \ \ \ \ \ \ \ \ \ \ \mp
\left.
{\cal Z}^{(N_{f}+2)}_{\nu=+1/2}(\{\mu_i\},\zeta,\om)
\ {\cal Z}^{(N_{f})}_{\nu=-1/2}(\{\mu_i\})
\right].
\label{factor3}
\eeqn

\noi\noindent
{\sc Proof}: In order to prove these two statements we will first
rephrase them in terms of linear combinations of QCD$_3$ partition
functions\footnote{Note that there is nothing new in the definition
(\ref{ZQCD3zw}) as compared with the original definition
(\ref{ZQCD3evenexpl}). In order to avoid confusion, we have
indicated explicitly where the unpaired masses should enter in the
individual blocks.}:
\beqn
{\cal Z}^{(2N_{f}+2)}_{{\rm QCD}_3}(\{\mu_i\},\zeta,\om)
&+&  {\cal Z}^{(2N_{f}+2)}_{{\rm QCD}_3}(\{\mu_i\},-\zeta,\om) \ = \
\label{factor3+}\\
&=& 2\pi^{N_f+1} \sqrt{\zeta\om}\
{\cal Z}^{(N_{f}+2)}_{\nu=-1/2}(\{\mu_i\},\zeta,\om)
\ {\cal Z}^{(N_{f})}_{\nu=+1/2}(\{\mu_i\}) \ ,\nn
\eeqn
and
\beqn
{\cal Z}^{(2N_{f}+2)}_{{\rm QCD}_3}(\{\mu_i\},\zeta,\om)
&-&  {\cal Z}^{(2N_{f}+2)}_{{\rm QCD}_3}(\{\mu_i\},-\zeta,\om) \ = \
\label{factor3-}\\
&=&-2 \pi^{N_f+1} \sqrt{\zeta\om}\
{\cal Z}^{(N_{f}+2)}_{\nu=+1/2}(\{\mu_i\},\zeta,\om)
\ {\cal Z}^{(N_{f})}_{\nu=-1/2}(\{\mu_i\}) \ .\nn
\eeqn
Clearly, when taking the sum or difference of these two equations we
find back the theorem eq. (\ref{factor3}). In the following we will
reduce eqs. (\ref{factor3+})  to identities between determinants
which will be then proven separately in
Appendix \ref{3-4dets}.

\noi
First of all, we note that, as in the proofs of the previous theorems
on the QCD$_3$ side, we can again reorder the determinant in terms of
hyperbolic functions.
After permuting the additional arguments
$\zeta$ and $\om$ we obtain
\beq
{\cal Z}^{(2N_{f}+2)}_{{\rm QCD}_3}(\{\mu_i\},\zeta,\om) \ =\
(-1)^{(N_f+1)(N_f+2)/2}2^{N_f+1}
\frac{\det {\matD}(\{\mu_i\},\zeta,\om)}
{\Delta(\{\mu_i\},\{-\mu_i\},\zeta,\om)} \ ,
\eeq
where we define the $(2N_f+2)\times(2N_f+2)$ determinant $\matD$
for an even number of flavors,
\beq
\det {\matD}(\{\mu_i\},\zeta,\om)\ =\
\det\left(
\begin{array}{ll}
\matC(\{\mu_i\})  & -\matS(\{\mu_i\})  \\
\matC(\{\mu_i\})  &\ \ \matS(\{\mu_i\})\\
\zeta^{j-1}\cosh^{(j-1)} (\zeta) &-\zeta^{j-1}\sinh^{(j-1)} (\zeta) \\
\om^{j-1}\cosh^{(j-1)} (\om) &-\om^{j-1}\sinh^{(j-1)} (\om)
\end{array}\right) \ .
\label{matDdef}
\eeq
In the next step, we write out the Vandermonde determinants
explicitly as in eq. (\ref{Deltarelodd}),
\beq
\Delta(\{\mu_i\},\{-\mu_i\},\pm\zeta,\om) \ = \ (-1)^{N_f(N_f+1)/2}
\ (\om\mp\zeta)
\prod_{f=1}^{N_f}2\mu_f
\ \Delta(\{\mu_i^2\})^2
\prod_{j=1}^{N_f}(\om^2-\mu_j^2)(\zeta^2-\mu_j^2)
\eeq
for the QCD$_3$ side, and
\beq
\Delta(\{\mu_i^2\},\zeta^2,\om^2)\Delta(\{\mu_i^2\})\ = \
(\om^2-\zeta^2)\ \Delta(\{\mu_i^2\})^2
\prod_{j=1}^{N_f}(\om^2-\mu_j^2)(\zeta^2-\mu_j^2)
\eeq
for the QCD$_4$ side. Putting everything together, we obtain from
eq. (\ref{factor3+})
\beq
(\om+\zeta)\det {\matD}(\{\mu_i\},\zeta,\om) \ +\
(\om-\zeta)\det {\matD}(\{\mu_i\},-\zeta,\om)
\ =\
(-1)^{N_f+1}2^{N_f+1}\det\matC(\{\mu_i\},\zeta,\om)\det\matS(\{\mu_i\})
\label{detrel+}
\eeq
and from eq. (\ref{factor3-})
\beq
(\om+\zeta)\det {\matD}(\{\mu_i\},\zeta,\om) \ -\
(\om-\zeta)\det {\matD}(\{\mu_i\},-\zeta,\om)
\ =\
(-1)^{N_f}2^{N_f+1}\det\matS(\{\mu_i\},\zeta,\om)\det\matC(\{\mu_i\})
\ . \label{detrel-}
\eeq
These two relations between determinants are derived in Lemma 1 (\ref{La1})
and Lemma 2 (\ref{La1}) of Appendix A, which completes our proof of
Theorem III.

\vspace{0.5cm}
\section{Applying the Replica Method: General Results}
\vspace*{0.3cm}

\noi\indent
In this section we will use the replica method to
derive general relations between the microscopic
spectral densities of QCD$_3$ and QCD$_4$.

\noi
Let us first consider QCD$_4$ with
$N_f$ fermions in the fundamental representation. We add  $n$
valence fermions with degenerate (rescaled) mass $\mu_v$ to this
theory. In such a theory, the resolvent, or
partially quenched chiral condensate, is defined as
\beq
\frac{\Sigma_{\nu}(\mu_v,\{\mu_i\})}{\Sigma} ~\equiv~
\lim_{n\to0}\frac{1}{n}\frac{\partial}{\partial\mu_{v}}
\ln{\cal Z}^{(N_{f}+n)}_{\nu}(\mu_v,\{\mu_i\}) ~.
\label{Sigmadef}
\eeq
The partially quenched chiral condensate in QCD$_3$ is defined similarly
(recalling that here the number of flavors is $2n$). The
microscopic spectral density of the Dirac operator is given by
the discontinuity of the resolvent across the imaginary axis \cite{OTV},
\beqn
\rho_S^{(\nu)}(\zeta;\{\mu_i\}) &=& \frac{1}{2\pi} {\mbox{\rm Disc}}
\left.\Sigma_{\nu}(\mu_v,\{\mu_i\})\right|_{\mu_v=i\zeta} \cr
&=& \frac{1}{2\pi}[\Sigma_{\nu}(i\zeta+\epsilon;\{\mu_i\}) -
\Sigma_\nu(i\zeta-\epsilon;\{\mu_i\})]
\label{rhodisc}
\eeqn
The analysis of Dirac operator spectra by means of a generating function
of the resolvent is not restricted to just the spectral density.
The $k$-point spectral
correlation functions can be computed from partially quenched chiral
susceptibilities $\chi(\mu_{v_1},\ldots,\mu_{v_k},\{\mu_i\})$. These
quantities can be derived from the partition function by extending it
with $kn$ valence fermions in $k$ mass-degenerate sets. Then the
$n \to 0$ limit is again taken after appropriate differentiations,
\beq
\chi_{\nu}^{(k)}(\mu_{v_1},\ldots,\mu_{v_k},\{\mu_i\}) ~\equiv~
\lim_{n\to 0}
\frac{1}{n^k}\frac{\partial}{\partial
\mu_{v_1}}\cdots\frac{\partial}{\partial \mu_{v_k}}\ln
{\cal Z}_{\nu}^{(N_{f}+kn)}(\mu_{v_{1}},\ldots,\mu_{v_{k}},\{\mu_i\}) ~.
\label{chidef}
\eeq
In a spectral representation,
\beq
\chi_{\nu}^{(k)}(\mu_{v_1},\ldots,\mu_{v_k},\{\mu_i\}) ~=~
\int_{-\infty}^{\infty}\! d\zeta_1\cdots d\zeta_k
\ \frac{\rho_{S,{\rm conn.}}^{(\nu)}(\zeta_1,\ldots,\zeta_k;\{\mu_i\})}{
(i\zeta_1 + \mu_{v_{1}})\cdots (i\zeta_k + \mu_{v_{k}})} ~,
\eeq
and we can again invert this relation to get the spectral correlation
function as the discontinuity across the imaginary mass axis,
\beq
\rho_{S,{\rm conn.}}^{(\nu)}(\zeta_1,\ldots,\zeta_k;\{\mu_i\}) ~=~
\frac{1}{(2\pi)^k} {\mbox{\rm Disc}}
\left.\chi_{\nu}^{(k)}(\mu_{v_{1}},\ldots,\mu_{v_{k}},\{\mu_i\})
\right|_{\mu_{v_{j}}=i\zeta_j} ~.
\label{chidisc}
\eeq
For instance, for the 2-point function the explicit relation reads
after using the symmetry of the spectrum around the origin
\beqn
\rho_{S,{\rm conn.}}^{(\nu)}(\zeta_1,\zeta_2;\{\mu_i\}) &=&
\frac{1}{4\pi^2}[\chi_{\nu}^{(2)}(i\zeta_1+\epsilon,i\zeta_2+\epsilon;
\{\mu_i\}) + \chi_{\nu}^{(2)}(i\zeta_1-\epsilon,i\zeta_2-\epsilon;
\{\mu_i\}) \cr &&
\ \ \ \ - \chi_{\nu}^{(2)}(i\zeta_1-\epsilon,i\zeta_2+\epsilon;
\{\mu_i\}) - \chi_{\nu}^{(2)}(i\zeta_1+\epsilon,i\zeta_2-\epsilon;
\{\mu_i\})] ~.
\eeqn
A similar formula
gives us $\rho_{S,{\rm conn.}}^{(\nu)}(\zeta_1,\ldots,\zeta_k;\{\mu_i\})$
for arbitrary $k$.

\subsection{The Microscopic Spectral Density}
\vspace*{0.3cm}

\noi
Let us first apply the relation (\ref{rhodisc}) to the partition
function identity
(\ref{factoreven}), adding to the $2N_f$ physical fermions
$n$ valence fermions of mass $\mu_v$ and $n$ valence fermions of
mass $-\mu_v$. The number $n$ is eventually taken to
zero. We then obtain a relation between the
microscopic spectral density of
the Dirac operator in 2+1 dimensions for $2N_f$ fermions pairwise grouped
with opposite mass and the corresponding                               
microscopic spectral densities in QCD$_4$ {\em evaluated at half-integer
topological charge},
\be
\rho_{{\rm QCD}_{3}}^{(2N_{f})}(\zeta;\{\mu_i\})&=&
\frac12\ \left[ \rho_{{\rm QCD}_{3}}^{(2N_{f})}(\zeta;\{\mu_i\}) +
\rho_{{\rm QCD}_{3}}^{(2N_{f})}(-\zeta;\{\mu_i\})\right]  \nn\\
 &=&
\frac{1}{2}\left[\rho_{{\rm QCD}_4}^{(N_{f},\nu=+1/2)}(\zeta;\{\mu_i\}) +
\rho_{{\rm QCD}_4}^{(N_{f},\nu=-1/2)}(\zeta;\{\mu_i\})\right].
\label{rho34even}
\ee
In the first equality we have used that the {\it average} spectral density is
an even function of $\zeta$.
Note that the $\mu$-independent factors in eqs. (\r{factoreven})
and (\r{factorodd}) are immaterial.
The same relation can be derived from the
Random Matrix Theory representation of the finite volume partition function
(see Appendix B).

\noi
For $2N_f+1$ flavors we only consider
the case ${\cal Z}_{{\rm QCD}_3^+}^{(2N_f+1)}$ for which the spectral density
is positive definite.
{}From the replica
limit of the resolvent applied to (\ref{factorodd})
we find
\beq
\rho_{{\rm QCD}_3^+}^{(2N_{f}+1)}(\zeta;\{\mu_i\},\mu) ~=~ \frac{1}{2}
\left[\rho_{{\rm QCD}_4}^{(N_{f},\nu=-1/2)}(\zeta;\{\mu_i\},\mu) +
\rho_{{\rm QCD}_4}^{(N_{f},\nu=+1/2)}(\zeta;\{\mu_i\})\right],
\label{rho34odd}
\eeq
where we again have used that the average spectral density is an
even function of $\zeta$.
For $\mu=0$ this relation can be simplified
by applying flavor-topology duality \cite{Jac},
\beq
{\cal Z}_{{\rm QCD}_3^+}^{(2N_f+1)}(0,\{\mu_i\})
~=~ \pi^{N_f}
\prod_{j=1}^{N_f}\frac{1}{\mu_j}
\left( {\cal Z}_{\nu=+1/2}^{(N_f)}(\{\mu_i\})\right)^2 ~.
\label{factoroddml+}
\eeq
Here, the prefactor gets modified compared to eq. (\ref{factorodd})
for the following reason.
If we define all partition functions normalized according to
eqs. (\ref{ZchUEexpl}) and (\ref{matBdef}) the flavor-topology shift
for general $\nu$ (see
eq. (\ref{FTfactornu})) yields the following factor:
\beq
\lim_{\mu\to 0}
\left(\frac{\pi\mu}{2}\right)^{1/2}
{\cal Z}_{\nu=-1/2}^{(N_f+1)}(\mu,\{\mu_i\})
~=~ \prod_{j=1}^{N_f}\frac{1}{\mu_j}\ {\cal Z}_{\nu=+1/2}^{(N_f)}(\{\mu_i\})~.
\label{FTfactor}
\eeq
Eq. (\ref{factoroddml+}) leads to
\beq
\rho_{{\rm QCD}_3}^{(2N_{f}+1)}(\zeta;\{\mu_i\},0) ~=~
\rho_{{\rm QCD}_4}^{(N_{f},\nu=+1/2)}(\zeta;\{\mu_i\}) ~,
\label{rho34oddml}
\eeq
a relation first discovered by Christiansen \cite{Jesper} in the context of
Random Matrix Theory. Here we see that this relation follows from the
effective field theory partition functions alone, after using the
replica method. Note how the ``topological'' $\nu/\mu$-term
(for $\nu = 1/2$)
nicely has
cancelled out on the right hand side.

\noi
The relations (\ref{rho34even}) and (\ref{rho34odd}) are our two main
results. They show that the microscopic spectral densities of QCD$_3$
follow directly from those of QCD$_4$ by analytic continuation in
$\nu$.

\noi
It is straightforward to check the relations
(\ref{rho34even}) and (\ref{rho34odd}) against the
well-known expressions for the {\em massless} microscopic spectral densities
for the two theories, as for example computed in
Random Matrix Theory \cite{SV,VZ},
\newpage
\beqn
\rho_{{\rm QCD}_3}^{(2N_{f})}(\zeta;\{\mu_i=0\}) &=&
\frac{\zeta}{4}\left[J_{N_{f}-1/2}(\zeta)^2
- J_{N_{f}+1/2}(\zeta)J_{N_{f}-3/2}(\zeta) \right. \cr
&& \left.+ \  J_{N_{f}+1/2}(\zeta)^2
- J_{N_{f}+3/2}(\zeta)J_{N_{f}-1/2}(\zeta)\right] ~,\cr
\rho_{{\rm QCD}_3}^{(2N_{f}+1)}(\zeta;\{\mu_i=0\},0) &=& \frac{\zeta}{2}\left[
J_{N_{f}+1/2}(\zeta)^2
- J_{N_{f}+3/2}(\zeta)J_{N_{f}-1/2}(\zeta)\right] ~,\cr
&& \cr
\rho_{{\rm QCD}_4}^{(N_{f},\nu)}(\zeta;\{\mu_i=0\}) &=& \frac{\zeta}{2}\left[
J_{N_{f}+\nu}(\zeta)^2
- J_{N_{f}+\nu+1}(\zeta)J_{N_{f}+\nu-1}(\zeta) \right]\ ~.
\label{rhos}
\eeqn
The relations derived above are indeed seen to be satisfied.

\noi
It is also instructive to write down the asymptotic large mass
expansion of the resolvent of the QCD$_3$ partition function from the known
asymptotic expansion of the resolvent for the QCD$_4$ partition function.
We only consider the case of an even number of massless flavors.
The starting point is the relation
(\ref{factoreven}) for $2N_f$ massless flavors and $2n$ valence flavors
with masses in pairs $\pm \mu_v$. Using flavor-topology duality \cite{Jac},
the QCD$_4$ partition function can be rewritten using
\be
\lim_{\mu\to
  0}\prod_{k=1}^{n}\left[\Gamma(\nu+k)\left(\frac{2}{\mu}\right)^{
\nu+k-1}\right]
{\cal Z}^{(N_f+n)}_\nu (\{\mu,\cdots,\mu,\mu_v\}) \ =\  \frac{
{\cal Z}^{(n)}_{N_f+\nu} (\{\mu_v\})}{  \mu_v^{n(\nu+N_f)}}
\ee
This relation is readily derived from the explicit form of the finite volume
partition functions \cite{Jac}.
{}From (\ref{factoreven}) we then obtain
\be
{\cal Z}_{{\rm QCD}_3}^{(2N_f + 2n)}(\{0,\cdots, 0, \mu_v\}) \ =\
\frac{\pi^{N_f+n}}{ \mu_v^{2nN_f}}
\ {\cal Z}^{( n)}_{\nu = N_f -1/2}(\{\mu_v\})
\ {\cal Z}^{(n)}_{\nu = N_f +1/2}(\{\mu_v\}).
\ee
After taking the replica limit we find the relation
\be
\frac{\Sigma_{{\rm QCD}_3}^{2N_f}(\mu_v)}\Sigma \ =\
\frac12 \left [
 \frac{\Sigma_{\nu=N_f-1/2}^{N_f=0}(\mu_v)}{\Sigma}
+\frac {\Sigma_{\nu=N_f  +1/2}^{N_f=0}(\mu_v)}{\Sigma}
\right ] -\frac{N_f}{\mu_v}.
\ee
The replica limit of the quenched QCD$_4$ chiral condensate
is given
by (for $ -\frac 32 \pi < {\rm arg}\ \mu_v < \frac \pi 2$) \cite{DV}
\be
\frac{\Sigma_\nu(\mu_v)}{\Sigma}
&=& 1 -\frac{i(-1)^{\nu}e^{-2\mu_v}}{2\mu_v} +
\frac{(4\nu^2-1)(1-i(-1)^{\nu}e^{-2 \mu_v})}{8\mu_v^2}
-\frac{i(-1)^{\nu}e^{-2\mu_v}(4\nu^2-1)(4\nu^2-9)}
{64\mu_v^3}\nonumber\\
&&+\ \frac{(4\nu^2-1)(4\nu^2-9)\left\lbrack
i(-1)^{\nu}e^{-2\mu_v}(19-4\nu^2)-6\right\rbrack}{3\times 256 \mu_v^4}
\ + \cdots\ .
\label{cond4}
\ee
We thus find the following large mass expansion for the resolvent of
QCD$_3$
\be
\frac {\Sigma_{{\rm QCD}_3}^{2N_f}(\mu_v)}\Sigma &=&
1 -\frac{N_f}{\mu_v} +
\frac{N_f^2}{2\mu_v^2}-\frac {N_f(N_f^2-1)}{4\mu_v^4} \nn\\
&&+\  e^{-2\mu_v} (-1)^{N_f}\left
[\frac{N_f}{2\mu_v^2} +\frac{N_f(N_f^2-1)}{2\mu_v^3}
+\frac{N_f(N_f^2-1)(N_f^2-3)}{4\mu_v^4} \right] +\cdots \ .
\label{QCD3resolvent}
\ee
For $N_f =1$ and $N_f =2$ all higher-order terms vanish. A replica calculation
thus provides us with the exact non-perturbative result.

\subsection{Higher Order Correlation Functions}
\vspace*{0.3cm}

\noi
Let us now apply  the relations between partition
functions in QCD$_3$ and QCD$_4$ to higher order correlation functions.
Because the valence masses occur in pairs $\pm \mu_k$ in the
QCD$_3$ partition function, differentiation with respect to the valence
quark masses does not exactly result in
the $k$-point spectral correlation functions. Rather we obtain nontrivial
relations between correlation functions with arguments that differ by a
minus sign.
As an example, let us look in more detail at the
two-point correlation function. We find
\be
&&\rho_{{\rm QCD}_3,{\rm conn.}}^{(2N_{f})}(\zeta_1,\zeta_2;\{\mu_i\})
\ +\ \rho_{{\rm QCD}_3,{\rm conn.}}^{(2N_{f})}(\zeta_1,-\zeta_2;\{\mu_i\})
\nn\\
+ &&\rho_{{\rm QCD}_3,{\rm conn.}}^{(2N_{f})}(-\zeta_1,\zeta_2;\{\mu_i\})
\ +\ \rho_{{\rm QCD}_3,{\rm conn.}}^{(2N_{f})}(-\zeta_1,-\zeta_2;\{\mu_i\})
\nonumber\\
= && \rho_{S,{\rm conn.}}^{(\nu=+1/2)}(\zeta_1,\zeta_2;\{\mu_i\})\ +\
\rho_{S,{\rm conn.}}^{(\nu=-1/2)}(\zeta_1,\zeta_2;\{\mu_i\}) \ .
\label{two-relation}
\ee
With connected two-point correlators given by the square of the kernel,
\be
\rho_{{\rm QCD}_3,{\rm conn.}}^{(2N_{f})}(\zeta_1,\zeta_2;\{\mu_i\}) &=&
-\left [ K_{\rm UE}^{(2N_{f})}(\{\mu_i\},\zeta_1,\zeta_2)\right ]^2 ~,
\nonumber \\
\rho_{S,{\rm conn.}}^{(\nu)}(\zeta_1,\zeta_2;\{\mu_i\}) &=&
-\left[ K_{\rm chUE}^{(N_{f},\nu)}(\{\mu_i\},\zeta^2_1,\zeta^2_2)\right]^2 ~,
\ee
the relation (\ref{two-relation}) is in complete agreement with result
from Random Matrix Theory obtained in Appendix B (see eq. (\ref{kernels1})).
Similar relations can be derived for higher
order correlation functions.

\noi
We stress that the above relations are totally general, independent of
any detailed evaluation of the $n \to 0$ limit of the partition
functions involved. The replica method thus provides us with
truly non-perturbative relations which
go beyond asymptotic expansions. Actually, in this case
the asymptotic expansions terminate, as we shall see below. 

Finally, a note of caution for using the replica method. In our proofs 
we have not excluded the possible appearance of terms that vanish      
for positive integer values of $n$ but are finite for $n\to 0$         
(for example, terms of the the form $\sin n\pi/n$). Because of the     
agreement with the exact results we know that such terms are absent.   
It is an interesting problem to understand why this is the case for    
an analytical continuation in the number of flavors whereas such       
terms appear, as we have seen in eq. (\ref{besselnu}),                  
when we perform an         
analytical continuation in the topological charge.                     

\setcounter{equation}{0}
\section{Virasoro Constraints and Spectral Sum Rules}
\vspace*{0.3cm}

\noi
In the next section
we will turn to explicit calculations based on the replica method. So far
there are no known techniques that have succeeded in taking the replica limit
$n\to 0$ beyond series expansions. The most obvious
expansion that comes to mind is a Taylor series in the mass. Such an
expansion was recently considered in the context of QCD$_4$ \cite{DS},
where it was found that the so-called de~Wit--'t~Hooft poles
prohibit the evaluation of the partially quenched chiral condensate
$\Sigma_{\nu}(\mu_v,\{\mu_i\})$ beyond a given order, depending on the
value of $N_f$ and $\nu$. In QCD$_4$, these de~Wit--'t~Hooft poles occur
at precisely the order for which the logarithmic terms appear in the expansion
\cite{DS,DV}. As a  particular consequence,
the small-mass expansion cannot be used
to derive spectral correlation functions. There
is simply no discontinuity across the imaginary mass axis up to the order
at which the replica method predicts the partially quenched chiral
susceptibilities.\footnote{Resummations of the series may hold a clue
to how this obstacle can be overcome; see section 4.2 of ref. \cite{DV}.}

\noi
As a preparation for the next section
we derive in this section
the  small-mass expansions of the partition functions
(\ref{ZQCD3even}) and (\ref{ZQCD3odd0}). Taking advantage
of the results obtained we also derive
the corresponding spectral sum rules, thus  generalizing
previous results. The small-mass expansions of
(\ref{ZQCD3even}) and (\ref{ZQCD3odd0}) can be
obtained by several different methods such as the character expansion
of \cite{IZ}\footnote{We have explicitly verified that the more general
expansion we shall provide below agrees up to fourth order with the
character expansion of \cite{IZ} for the case Tr$({\cal M})=0$.}.
We can also use the known results
for the effective partition function of QCD$_4$ as a generator for
unitary integrals (as the effective QCD$_4$ partition function is simply
an external field integral over the group U($N_f$)). However,
the most elegant and efficient  way
to push the small mass
expansion to arbitrarily high orders is probably through the
use of Virasoro constraints. The appearance of Virasoro
constraints in unitary integrals was first noticed in
\cite{GN} for the case of large masses. It was
later generalized  for the case of small masses in \cite{moro}.
In both works, only the one-link integral (\ref{ZchUE})
at zero topological charge ($\nu=0$) was studied, although in the
small-mass expansion the flavor-topology duality readily gives
the analytical continuation to $\nu \neq 0$.

\noi
For the  derivation of  Virasoro constraints for QCD$_3$ it is convenient
to start with the following more general partition function:
\beq
{\cal Z}_{q}^{(N)}(M) = \int dU \exp
\left[ {\rm Tr}MUI_qU^{\dagger}\right] ~,
\label{ZQ}
\eeq
where $M$ is an arbitrary $N\times N$ matrix
and $I_q$ is such that $\, I_q^2=${\bf 1}$_{N\times N} \,$
and $\, {\rm Tr}I_q=q\, $. In particular, the even-flavor partition
function of eq. (\ref{ZQCD3even}) is obviously of this kind (with $q=0$
and $M = V\Sigma {\cal M}$),
and it turns out that in the small-mass expansion other detailed
properties of $I_q$ are irrelevant.
Thus, in the small-mass expansion the partition functions (\ref{ZQ}) with
$q=0,\pm 1$ are the building blocks for the QCD$_3$ partition
functions given by (\ref{ZQCD3even}), (\ref{ZQCD3odd0})
and (\ref{ZQCD3odd1}). It was noted in ref. \cite{DS} that the
finite-volume effective field theory partition functions satisfy the
same type of differential equation, whether in 3 or 4 dimensions.
Indeed, from the properties of $I_q$ we find
\beqn
\left( \frac{\partial^2}{\partial M_{ca} \partial M_{ab}} - \delta_{cb}
\right){\cal Z}_{q}^{(N)} \, &=& \, 0 \label{d2m}\\
\delta^{ab}\frac{\partial {\cal Z}_{q}^{(N)}}
{\partial M_{ab}}\, &=&\, q\, {\cal Z}_{q}^{(N)}.
 \label{d1m}
\eeqn
Due to the unitary invariance of the measure  it is clear that
the partition function, (\ref{ZQ}),
depends only on the traces of the
matrix $M$.
Thus, following \cite{moro}, we can change variables from
the matrix elements $M_{ab}$ to
\beq
t_k \, = \, \frac 1k {\rm Tr}M^k \quad , \quad k\ge 1.
\eeq
Clearly, for finite $N$ not all $t_k$ are independent.
Using the chain rule, the differential equations
(\ref{d2m}) and (\ref{d1m}) can be rewritten as
\beqn
\sum_{s\ge 2}M_{bc}^{s-2}\left[{\cal L}_s -
\delta_{s,2}\right]{\cal Z}_{q}^{(N)}\,
&=& \, 0, \\
({\cal L}_1 - q ){\cal Z}_{q}^{(N)} \, &=& \, 0\, ,
\eeqn
where the operators ${\cal L}_s$ are defined by
\beq
{\cal L}_s\, = \, \sum_{j=1}^{s-1}\partial_j\partial_{s-j}
+\sum_{j\ge 1}jt_j\partial_{j+s} + N\partial_s, \qquad \, s\ge 1.
\eeq
We use the notation $\partial_j =\partial/\partial t_j $, and our  convention
is that for $s=1$ the first term in ${\cal L}_s $ is absent.
The ${\cal L}_s$  satisfy the Virasoro
algebra\footnote{A ``classical'' Virasoro algebra, since
there is no central charge. It is also known as a Witt algebra.}
\be
\left[ {\cal L}_r , {\cal L}_s \right] = (r-s)
{\cal L}_{r+s}.
\ee
A sufficient condition for the solution of the differential equations
(\ref{d2m}) and (\ref{d1m}) is thus given by so-called Virasoro
constraints:
\beq
{\cal L}_s {\cal Z}_{q}^{(N)} \,
= \, (\delta_{s,2}+q\delta_{s,1}) {\cal Z}_{q}^{(N)} \, ,  \quad s\ge 1 .
\eeq
They differ from the Virasoro constraints for
QCD$_4$  by the presence of the term $\delta_{s,2} $ and the parameter $q$
which takes the value $q=1$ in the QCD$_4$ case.

\noi
We normalize the partition function by
${\cal Z}_{q}^{(N)}(t_k=0)=1$. The Virasoro constraints
can then be solved iteratively order by order in the masses by
letting $t_k=0$ after acting with the differential operators.
It is instructive to consider the solution of the lowest few orders.
Using the Ansatz,
\beq
{\cal Z}_{q}^{(N)} = 1 + at_1 + b_1 t_1^2 + b_2 t_2 + {\cal O}(\mu^3) ~,
\eeq
we deduce from $
{\cal L}_1 {\cal Z}_{q}^{(N)}=q{\cal Z}_{q}^{(N)} $
at $t_k=0$,
\beq
N a \, = \, q.
\eeq
At order $\mu^2$ we can use
$\partial_1  {\cal L}_1 {\cal Z}^{(N)}_q = q \partial_1 {\cal Z}^{(N)}_q$
and ${\cal L}_2{\cal Z}_{q}^{(N)}={\cal Z}_{q}^{(N)} $
at $t_k=0$ to  obtain the equations,
\beqn
2Nb_1 + b_2 &=& q a \, = \,\frac{q^2}{N} \nn,\\
2b_1 + Nb_2 &=& 1 .
\eeqn
The solution is  given by
$b_2=(N^2-q^2)/(N(N^2-1)) $ and  $b_1=(q^2-1)/(2(N^2-1)) $.
Proceeding like this, we have derived the small-mass expansion
of ${\cal Z}_q^{(N)}$ up to order $\mu^6$, and the expansion can easily
be pushed higher. We quote here the ${\cal O}(\mu^4)$ result,
which is given by
\beqn
{\cal Z}_{q}^{(N)}&=&
\, 1 + \frac{q}{N} {\rm Tr}M +
\frac{(N^2-q^2){\rm Tr}M^2}{2N(N^2-1)} +\frac{(q^2-1)({\rm Tr}M)^2}{2(N^2-1)}
-\frac{2q(N^2-q^2)}{3N(N^2-1)(N^2-4)}{\rm Tr}M^3
 \,  \nn\\
&+&\frac{q(N^2-q^2)}{2(N^2-1)(N^2-4)}{\rm Tr}M{\rm Tr}M^2
+\frac{q}{6N}\left[\frac{N^2(q^2-3)+4-2q^2}{(N^2-1)(N^2-4)}\right]
({\rm Tr}M)^3  \nn\\
&+&\frac{(N^2-q^2)(5q^2+4-N^2)}{4N(N^2-1)(N^2-4)(N^2-9)}{\rm Tr}M^4
+ \frac{(N^2-q^2)(3N^2+3q^2-2N^2q^2-12)}{3N^2(N^2-1)
(N^2-4)(N^2-9)}{\rm Tr}M{\rm Tr}M^3 \nn\\
&+&
\frac{(N^2-q^2)(24-10N^2+N^4-6q^2-N^2q^2)}{8N^2(N^2-1)
(N^2-4)(N^2-9)}({\rm Tr}M^2)^2 +
 \frac{(N^2-q^2)(q^2-1)}{4N(N^2-1)(N^2-9)}({\rm Tr}M)^2{\rm Tr}M^2 \nn\\
&+& \frac{q^2\left[N^4(q^2-6)-8N^2(q^2-5)+6(q^2-4)\right] +3N^2(N^2-4)}
{24N^2(N^2-1)(N^2-4)(N^2-9)}({\rm Tr}M)^4 \, + \, {\cal O}(\mu^5) .
\label{m6}
\eeqn
In Appendix C we give the expansion up to order $\mu^6$.
At order $\mu^5$ we have found simple poles\footnote{We ignore the poles at
unphysical negative values of $N$, which are completely irrelevant
here.} at $N^2 = 0,1,4,9$ and 16,
while at order $\mu^6$, besides the simple poles at
$N^2 = 0,1,4,9,16$ and 25, a double pole appears at
$N=1$, in agreement with the
calculation of unitary integrals
in reference \cite{SAM}. We can check that our small-mass expansion
of eqs. (\ref{m6}), (\ref{z5}) and (\ref{ABCD}) is correct by considering
several special cases. For instance, if we take
the mass matrix proportional to the identity
$M=m \,$ {\bf 1}$_{N\times N}$ all poles cancel,
and, according to
our normalization, we should obtain ${\cal Z}_q=e^{mq}$. Its
small-mass expansion is indeed in agreement with our results.
This is one non-trivial check, for an arbitrary number of flavors
and arbitrary value of $q$. We get
another simple but non-trivial check for different masses by
comparing our result
with the small-mass expansion of the exact result
\cite{IZ} for
the partition function ${\cal Z}_q^{(N)}$ at fixed value of
the number of flavors $N$. For example, for
$N=2$ with mass matrix $M={\rm diag}(\mu_1,\mu_2)$ and $q=0$,
the partition function is given by
\beq
{\cal Z}_{{\rm QCD3}}^{(2)}(\mu_1,\mu_2)
=\frac{\sinh (\mu_1-\mu_2)}{\mu_1-\mu_2}.
\eeq
Expanding this expression for small masses,
and comparing with  eq. (\ref{m6}) and the
corresponding expressions in Appendix C, we find perfect agreement.

\noi
It is important to realize that all the poles at integer values of $N$
{\em cancel exactly} once we specify whether $N$ is even $(q=0)$ or
odd $(q=1)$. Thus
neither the even-$N$ nor the odd-$N$ partition functions have any
physical poles at all.

\noi
We have also checked that the expansion (\ref{m6}) is consistent with
the expansion obtained from eq. (\ref{factoreven}) and the known
small-mass expansion of the QCD$_4$ partition function from ref. \cite{DS}
for Tr$({\cal M})=0$. It is instructive to follow the
way the coefficients of the expansion of the QCD$_4$ partition functions at
$\nu = \pm 1/2$ combine to yield the corresponding coefficients for QCD$_3$
as in eq. (\ref{m6}). One observes that the appearance of the
$\nu=\pm 1/2$ factor on the right hand side of eq. (\ref{factoreven})
is related to
the {\em doubling} in number of flavors on the left hand side.
As a simple example, let us consider
\beq
{\cal Z}^{(N_{f})}_{\nu}(\mu) ~=~ \mu^{N_{f}\nu}\left(1 + \frac{N_{f}\mu^2}
{4(N_f + \nu)} + \ldots\right) ~,
\eeq
for $N_f$ equal-mass flavors. This gives
\beqn
{\cal Z}^{(N_{f})}_{\nu=1/2}(\mu){\cal Z}^{(N_{f})}_{\nu=-1/2}(\mu)
&=& 1 + \frac{N_f}{2}\mu^2\left(\frac{1}{2N_f+1} + \frac{1}{2N_f-1}\right) +
\ldots \cr
&=& 1 + \frac{1}{2}\frac{(2N_f)^2}{(2N_f)^2-1}\mu^2 + \ldots ~,
\eeqn
which coincides with our expansion (\ref{m6}) to this order (in agreement
with equation (\ref{factoreven}) in the present normalization, where
$N=2N_f$ and $q=0$.).

\noi
For an {\it even} number of eigenvalues of the Dirac operator
the effective finite-volume partition functions for QCD$_3$, eqs.
(\ref{ZQCD3even}) and (\ref{ZQCD3odd0}),
yield the same small-mass expansions as
${\cal Z}_{{\rm QCD}_3}^{(N)}({\cal M})=(1/2)({\cal Z}_q^{(N)}
({\cal M}V\Sigma) +{\cal Z}_{-q}^{(N)}({\cal M}V\Sigma))$
where $q=0,1$ for an even and odd number of flavors, respectively.
Comparing the expansion
(\ref{m6}) with the small mass expansion of the QCD$_3$ partition function,
$\langle \prod_{a=1}^N \det (D +m_a) \rangle_{\rm Yang-Mills}$,
we can read off the massless spectral sum rules for the QCD$_3$ Dirac operator
\newpage
\beqn
\left\langle\sum_n \frac{1}{\zeta_n^2}\right\rangle \, &=& \,
\frac{N^2-q^2}{N(N^2-1)},\nn\\
 \left\langle\left(\sum_n \frac{1}{\zeta_n}\right)^2 \right\rangle
\, &=& \, \frac{1-q^2}{N^2-1}, \nn\\
\left\langle\sum_n \frac{1}{\zeta_n^4}\right\rangle \, &=& \,
\frac{(N^2-q^2)(N^2-5q^2-4)}{N(N^2-1)(N^2-4)(N^2-9)}, \nn\\
\left\langle\sum_n \frac{1}{\zeta_n^2} \left(\sum_m\frac{1}{\zeta_m}\right)^2
\right\rangle  \, &=& \,  \frac{(N^2-q^2)(1-q^2)}{N(N^2-1)(N^2-9)},
 \nn\\
\left\langle\left(\sum_n \frac{1}{\zeta_n}\right)^4\right\rangle
&=&\, \frac{q^2\left[N^4(q^2-6)-8N^2(q^2-5)+6(q^2-4)\right] +3N^2(N^2-4)}
{N^2(N^2-1)(N^2-4)(N^2-9)},   \nn\\
\left\langle\sum_n \frac{1}{\zeta_n} \sum_m\frac{1}{\zeta_m^3} \right\rangle
\, &=& \, \frac{(N^2-q^2)(3N^2+3q^2-2N^2q^2-12)}{N^2(N^2-1)
(N^2-4)(N^2-9)}, \nn\\
\left\langle\left(\sum_n \frac{1}{\zeta_n^2}\right)^2\right\rangle
\, &=& \, \frac{(N^2-q^2)(24-10N^2+N^4-6q^2-N^2q^2)}{N^2(N^2-1)
(N^2-4)(N^2-9)}.
\label{sumrules}
\eeqn
The sum rules obtained at order $\mu^6$ are given in the Appendix B.
It should be noticed that some sum rules given
in (\ref{sumrules}) and (\ref{sumrules6}) could not be derived
had we assumed ${\rm Tr}M^{2k+1}=0$ from the start.
The point is that we only deduce sum rules for the massless theory.
The quark masses only act as sources which should
be as general as possible; otherwise
some sum rules cannot be derived due to the
vanishing of the corresponding source term.

\noi
The first two massless sum rules agree with what was found in the original
paper \cite{VZ} for even and
odd flavors.
We have also explicitly checked that the ``diagonal'' sum rules
in (\ref{sumrules}) and (\ref{sumrules6})
agree with the results obtained from Random Matrix
Theory by integrating the first two formulas of (\ref{rhos}).
For the higher point functions a direct comparison
with higher spectral correlation functions obtained
from Random matrix theory becomes quite cumbersome. We have
instead generated general relations among sum rules
directly from the corresponding Random Matrix Theory via
Schwinger-Dyson equations. For instance,
from the identity
\beq
\int_{-\infty}^{\infty}
\prod_{j=1}^{{\cal N}}d\lambda_j \, \sum_{l}\partial_l\left[
\frac 1{\lambda_l}\left(\sum_k\frac 1{\lambda_k}
\right)^a\Delta^2 (\{\lambda_i\})
\prod_{p=1}^{{\cal N}}\left( \lambda_p^{2N_f}e^{-\frac {{{\cal N}}\Sigma^2}2
\lambda_p^2}\right)\right]\, =\, 0 \quad , \label{s-dyson}
\eeq
we derive, after the rescaling $\zeta_k={{\cal N}}\Sigma \lambda_k$,
\beq
\left\langle \left(\sum_k\frac 1{\zeta_k}\right)^{a+2} \right\rangle
\, + \, a\, \left\langle \sum_k\frac 1{\zeta_k^3}
\left(\sum_l\frac 1{\zeta_l}\right)^{a-1} \right\rangle
\, +\, \left\langle \left(\sum_k\frac 1{\zeta_k}\right)^{a} \right\rangle
\, = \, 2N_f\, \left\langle \sum_k\frac 1{\zeta_k^2}
\left(\sum_l\frac 1{\zeta_l}\right)^{a} \right\rangle\, .
\nn \\
\label{ward}
\eeq
We have checked that relation (\ref{ward}) for $a=0,2,4$,
as well as other similarly derived relations,
are in agreement with (\ref{sumrules}) and (\ref{sumrules6}).

\noi
For even ${{\cal N}}$,
the spectral sum rules that are odd under $\zeta_k \to - \zeta_k$
vanish identically as in the case of, $e.g.$,
$\langle \sum _k (1/\zeta_k) \rangle =0 $. This is a consequence
of the symmetry
${\cal Z}_{{\rm QCD}_3}^{(N)}(-M)={\cal Z}_{{\rm QCD}_3}^{(N)}(M)$.
This symmetry is clear for (\ref{ZQCD3odd0}), and
for (\ref{ZQCD3even}) it follows from the following argument.
In a small-mass expansion of (\ref{ZQCD3odd0})  for each
power of the matrix $M$ we  have the same
power of the matrix $\Gamma_5$
and the trace of odd powers of the mass matrix
must be accompanied by the trace of odd powers
of $\Gamma_5$ which vanishes identically.

\setcounter{equation}{0}
\section{The Replica Limit of the QCD$_3$ Finite Volume Partition Function}
\vspace*{0.3cm}

\noi
In this section, starting from the replica limit
of the small and large-mass expansions
of the QCD$_3$ finite volume partition functions,
we obtain the corresponding expansions of the
partially quenched chiral condensate.

\subsection{Small-Mass Expansion}
\vspace*{0.3cm}

\noi
Once we have the expansion of the QCD$_3$ partition function, it
should in principle be a simple matter to use the replica method to
derive the partially quenched chiral condensate. However, in the case
of the small-mass expansion this is not at all the case. The reason is
the apparent poles in the expansion. As mentioned above, whether $N$
is even or odd, all these poles cancel out exactly in the expansion
of the partition function itself. But with the replica method we are
forced to consider non-integer values of $N= 2N_f+2n$, where $n$ is
taken to zero. Then these so-called de Wit--'t Hooft poles \cite{D't}
are significant: the analytic continuation cannot be performed across
such poles. For larger and larger values of $N_f$ the first pole is
pushed to higher and higher order, but there will always be a finite
order beyond which we cannot construct $\Sigma(\mu_v;\{\mu_i\})$ from a
small-mass expansion using the replica method.
This issue was discussed at length in the case of
QCD$_4$ \cite{DS}, and we shall therefore be brief here.

\noi
For simplicity, let us restrict ourselves to the case of
massless physical fermions; the general case can be derived
in a completely analogous manner. For a mass matrix ${\cal M}$ with
$2N_f$ massless flavors and $n$ replica
flavors of mass $+\mu_v$ paired with $n$
replicas of mass $-\mu_v$ we have
$N=2n+2N_f$ and,
\beqn
{\rm Tr}({\cal M}^k) & = & 0 \, , \quad
{\rm for } \quad {\rm odd}\,\,  k,\nn\\
{\rm Tr}({\cal M}^k) & = & 2n\mu_v^k \, , \quad  {\rm for} \quad
{\rm even} \,\, k.
\eeqn
Using (\ref{Sigmadef}), we formally find
from the expansion of the partition function
(\ref{m6}) and (\ref{ABCD}):
\beqn
\frac{\Sigma(\mu_v)}{\Sigma}\, &=& \,
\frac{(4N_f^2-q^2)}{2N_f(4N_f^2-1)}\, \mu_v +
\frac{(4N_f^2-q^2)(5q^2+4-4N_f^2)}{2N_f(4N_f^2-1)(4N_f^2-4)
(4N_f^2-9)}\,\mu_v^3\nn\\
&+&\, \frac{2(4N_f^2-q^2)\left[(4N_f^2-4)(4N_f^2-16)+ 21q^4-14q^2 4N_f^2+140q^2
\right]}{2N_f(4N_f^2-1)(4N_f^2-4)(4N_f^2-9)(4N_f^2-16)(4N_f^2-25)}\,\mu_v^5 +
\cdots \quad .
\label{condreplica}
\eeqn
Here we have $q=0$ for integer $N_f$ and $q=1$ for half-integer
$N_f$, and we always
consider the case of an even number of Dirac eigenvalues.
As stressed above, the above expansion is only meaningful up to the
order at which a pole has to be crossed in order to take the replica limit.
The result (\ref{condreplica}) can now be compared with the expansion
of the result obtained from Random Matrix
Theory \cite{VZ}\footnote{For positive replica mass. The condensate is odd
in the mass, and appropriate absolute value signs have to be inserted
for $\mu_v < 0$; for example $\mu_v^2 \to \mu_v|\mu_v|$, etc..}:
\beqn
\frac{\Sigma_{{\rm RMT}}(\mu_v)}{\Sigma}(2N_f=1)&=&
\mu_v - \frac 23 \mu_v^2 +
\frac 13 \mu_v^3 - \frac{2}{15}\mu_v^4 + \frac{2}{45}\mu_v^5 + \cdots\ ,\nn\\
\frac{\Sigma_{{\rm RMT}}(\mu_v)}{\Sigma}(2N_f=2)&=&
\frac 23 \mu_v -
\frac 13 \mu_v^2 + \frac{2}{15}\mu_v^3 - \frac{2}{45}\mu_v^4 +
\frac{4}{315}\mu_v^5 + \cdots \ ,\nn\\
\frac{\Sigma_{{\rm RMT}}(\mu_v)}{\Sigma}(2N_f=3)&=&
\frac 13 \mu_v -
\frac 1{15} \mu_v^3 + \frac{2}{45}\mu_v^4 - \frac{2}{105}\mu_v^5
+ \cdots \ ,\nn\\
\frac{\Sigma_{{\rm RMT}}(\mu_v)}{\Sigma}(2N_f=4)&=&
\frac 4{15} \mu_v -
\frac 4{105} \mu_v^3 + \frac{1}{45}\mu_v^4 - \frac{8}{945}\mu_v^5
+ \cdots \ ,\nn\\
\frac{\Sigma_{{\rm RMT}}(\mu_v)}{\Sigma}(2N_f=5)&=&
\frac 1{5} \mu_v -
\frac 1{105} \mu_v^3 +  \frac{2}{945}\mu_v^5
+ \cdots \ .
\eeqn
Comparing with the expression (\ref{condreplica})
obtained from the replica method we indeed find exact
agreement precisely up to the order at which the replica method
no longer can be applied, as a de Wit--'t Hooft pole has to be crossed.
We nevertheless notice that, remarkably, {\em all} odd-order
coefficients agree with those of Random Matrix Theory if one
proceeds beyond the order at which the replica method ceases to
be valid. This is perhaps less mysterious if one considers the fact
that as $N_f \to \infty$ the replica expansion becomes exact to
all orders (and this exact expansion has just odd powers).
Because the replica method applied to the small-mass expansion only
gives us the (correct) series up to the first de Wit-`t Hooft pole,
we can only extract information about the microscopic spectral
density of the Dirac operator up to that power in the mass
which is free from such poles.

\subsection{Large-Mass Expansion}
\vspace*{0.3cm}

\noi
In this subsection we use the replica method to
calculate the large-mass
expansion of the partially quenched chiral condensate. Because
the expansion terminates, we are able to derive the exact
microscopic spectral density of the Dirac operator
in QCD$_3$ {\em without} relying on the assumption that it is
the universal microscopic spectral density of Random Matrix Theory.
 We perform this calculation  for one and two massless
flavors by a direct replica calculation  of the partition
function (\ref{ZQCD3even}).

\noi
We evaluate the partition function (\ref{ZQCD3even}) with
$U$ an $2(N_f+n)\times 2(N_f+n)$ unitary matrix.
The mass matrix is taken to be diagonal
with elements $M_{k+N_f+n} = - M_{k}$.
Ultimately, we are interested in the case
$M_1=\cdots = M_{N_f} =0$,
and the remaining replica masses equal to $\pm \mu$.
The integral (\ref{ZQCD3even})
can be rewritten as \cite{Vzirn},
\be
{\cal Z}_{{\rm QCD}_3}^{(2N_f+2n)} =
\int_{-1}^1 d \lambda_1 \cdots \int_{-1}^1 d \lambda_{N_f+n}
dU_1 dU_2 \Delta^2(\Lambda) e^{{\rm Tr} M_+
(U_1 \Lambda U_1^\dagger +U_2 \Lambda
U_2^\dagger)},
\ee
where $U_1$ and $U_2$ are $(N_f+n) \times (N_f+n)$ unitary matrices, $M_+$
is a diagonal matrix with diagonal elements given by $M_{+\, kk} =
M_{kk}$. The diagonal  matrix $\Lambda$ contains the integration
variables as diagonal entries. The Vandermonde determinant is denoted
by $\Delta(\Lambda) = \prod_{k>l} (\lambda_k -\lambda_l)$. The
integrals over $U_1$ and $U_2$ are familiar Itzykson-Zuber integrals.
The partition function is thus given by
\be
{\cal Z}_{{\rm QCD}_3}^{(2N_f+2n)} =
 \int_{-1}^1 d \lambda_1 \cdots \int_{-1}^1 d \lambda_{N_f+n}
\frac{{\det}^2 e^{M_k \lambda_l}}{\Delta^2(M)}.
\label{zvz}
\ee
In the limit of $N_f$ massless flavors and $n  $  quark masses equal
to $\mu$, the
determinant can be rewritten as
\be
\frac{\det e^{M_k \lambda_l}}{\Delta(M)} =
\frac{1}
{\mu^{N_f n}\prod_{k=1}^{N_f-1} k!\prod_{l=1}^{n-1} l!}
\left |
 \begin{array}{ccc} 1 & \cdots & 1 \\
\lambda_1 & \cdots & \lambda_{N_f+n} \\
                          \vdots &                 &\vdots \\
                         \lambda_1^{N_f-1} & \cdots &
\lambda_{N_f+n}^{N_f-1}\\
    e^{\mu\lambda_1} & \cdots& e^{\mu\lambda_{N_f+n}}\\
                           \lambda_1e^{\mu\lambda_1} & \cdots &
                \lambda_{N_f+n}e^{\mu\lambda_{N_f+n}} \\
                          \vdots &                 &\vdots \\
                         \lambda_1^{n-1}e^{\mu\lambda_1} & \cdots &
\lambda_{N_f+n}^{n-1}  e^{\mu \lambda_{N_f+n}}
\end{array} \right |.
\label{detter}
\ee
In general, this expression is rather complicated. However, the leading
term in an expansion in $1/\mu$ can be obtained easily. We will first derive
this term and then we will calculate the complete expansion for $N_f=1$ and
$N_f=2$.

\noi
There are two types of integrations: those that are accompanied by
an exponential $\exp \mu\lambda_k$, and those that are not. The second
type of integrals correspond to zero modes in a saddle point expansion
and have to be evaluated exactly. The remaining integrals are evaluated
perturbatively by writing
\be
\int_{-1}^1 d \lambda_k \cdots = \int_{-\infty}^1 d \lambda_k \cdots
- \int_{-\infty}^{-1} d \lambda_k \cdots \, .
\label{split}
\ee
and putting $\lambda_k = 1-x_k/\mu$ in the integrals over the segment
$\langle -\infty, 1]$, and  $\lambda_k = -1-x_k/\mu$ in the integrals
over the segment $\langle -\infty, -1]$. In both cases
the integral over
the $x_k$ ranges over the segment $[0,\infty\rangle$.
Denoting  by $p$ the number of integrals with $-1$ as upper limit,
the leading order contribution is obtained by keeping only the $p=0$ terms.
The leading order
term is then given by the diagonal term in the expansion of the square of
the determinant. All terms are the same and we thus find
\be
{\cal Z}_{{\rm QCD}_3}^{(2N_f+2n)} =
&& \frac{e^{2\mu n}}{\mu^{n^2+2N_f n}}
\int_{-1}^1 d \lambda_1 \cdots \int_{-1}^1 d \lambda_{N_f}
\Delta^2(\lambda_1, \cdots, \lambda_{N_f})\nonumber \\
&&\times \int_{0}^\infty d x_{N_f+1} \cdots \int_{0}^\infty d x_{N_f+n}
e^{-2 x_{N_f+1} - \cdots -2x_{N_f+n}} \Delta^2(x_{N_f+1},
\cdots, x_{N_f+n}) .
\ee
We recognize the partition function for the Laguerre ensemble which
can easily be evaluated by means of the orthogonal polynomial method.
However, we do not need this constant to find its leading order
large-$\mu$ contribution to the resolvent.
 In the replica limit $n \to 0$ it is simply given by
\be
{\cal Z}_{{\rm QCD}_3}^{(2N_f+2n,p=0)} =
 \frac{e^{2n \mu}}{\mu^{2n N_f}},
\ee
so that the $\mu$-dependence of the resolvent becomes
\be
\Sigma(\mu) = 1 - \frac{2N_f}\mu + O(1/\mu^2).
\ee

\noi
Let us return again to the group integral (\ref{ZQCD3even}).
For $p \ne 0$ the integrand is invariant under a $[U(n)/U(n-p_1)
\times U(p_1)]\times[U(n)/U(n-p_2)
\times U(p_2)]$ submanifold of the coset. The integral over this
manifold has to be performed exactly. Its volume is $\sim n^{p_1+p_2}$.
In the replica limit of the resolvent only terms with $p=p_1+p_2 = 0, \,
1$ contribute\footnote{It is important to stress that
$0\le p_1\, , \, p_2\le n $ and the reader may wonder
how can we make sense of sums over all saddle points
in the limit  $n\to 0 $. What we do in this section is equivalent      
to a presciption given in \cite{Mezard,lerner}.                        
This procedure has several caveats and                                 
is not mathematically rigorous \cite{Z1}, but in all cases where       
it has been applied \cite{Mezard,lerner,DV} it gives the correct result.}.  

\noi
Next, we evaluate the integral (\ref{zvz}) for $N_f=1$.
We expand this determinant with respect to its first row.
Using the permutation symmetry of the integration variables the square
of this expansion can be reduced to only two different contributions.
The partition function thus simplifies to (up to an overall constant)
\be
{\cal Z}_{{\rm QCD}_3}^{(2+2n)}
 = \frac 1{\mu^{2n}} \frac 1{(\prod_{k=1}^{n-1} k!)^2}
&&\int_{-1}^1 d \lambda_1 \cdots \int_{-1}^1 d \lambda_{n+1}
\biggl [
(n+1)e^{\mu(2\lambda_2 + \cdots +2\lambda_{n+1})} \Delta^2(\lambda_2,
\cdots, \lambda_{n+1})
\biggr .
\nonumber \\&& \!\!\!\!\biggl .
- n(n+1) e^{\mu(\lambda_1 +\lambda_2 + 2\lambda_3
+ \cdots + 2\lambda_{n+1})} \Delta(\lambda_1,\lambda_3,\cdots,
\lambda_{n+1})
\Delta(\lambda_2,\lambda_3,\cdots, \lambda_{n+1})
 \biggr ].\nonumber \\
\ee
We will expand this partition function
in powers of $1/\mu$. The first term in the integrand does not depend
on $\lambda_1$ and therefore the integral over $\lambda_1$ of this term
has to be performed exactly. It simply gives a factor 2. The remaining
integrals are evolved perturbatively. We again split the integrals
according to (\ref{split}) and put $\lambda_k =  \pm 1 -
x_k/\mu$ depending on the upper boundary of the integral.
For asymptotically large $\mu$, the integration range of the
$x_k$ is thus given by $[0, \infty \rangle$.
In the first term the only nonvanishing possibility is $p_1 =p_2 = 0$.
The term with $p_1 =p_2 =1 $ is of order $n^2$ and does not
contribute in the replica limit. However, in the second term we have
the possibilities $p_1 = 1, \, p_2 =0$ and $p_1 = 0, \, p_2 =1$
in addition the possibility $p_1 =p_2 =0$. All other contributions
vanish in the replica limit.
We thus find that the asymptotic expansion of the QCD$_3$ partition
function is given by
\be
{\cal Z}_{{\rm QCD}_3}^{(2+2n)}
  &=& \frac {e^{2n \mu}}
{\mu^{n^2 +2n} (\prod_{k=1}^{n-1} k!)^2}\biggl [
2(n+1)\int_{0}^\infty d x_2 \cdots \int_{0}^\infty d x_{n+1}
e^{-2 x_2 - \cdots -2x_{n+1}} \Delta^2(x_2,
\cdots, x_{n+1}) \biggl .\nonumber \\
&-&  \frac {n(n+1)} \mu \int_{0}^\infty\! d x_1
\cdot\cdot \int_{0}^\infty\! d x_{n+1}
e^{-x_1-x_2-2 x_3 - \cdots -2x_{n+1}}
\Delta(x_1,x_3,\cdots, x_{n+1})
\Delta(x_2,x_3,\cdot\,\cdot, x_{n+1})   \nonumber \\
&+&
\frac {2n(n+1)\mu^{n -1} e^{-2\mu}} \mu \int_{0}^\infty d x_1
\cdots \int_{0}^\infty d x_{n+1}
e^{-x_1-x_2-2 x_3 - \cdots -2x_{n+1}} \nonumber \\ \biggl . &&\qquad\times
\Delta(2 + \frac {x_1}\mu,x_3,\cdots, x_{n+1})
\Delta(x_2,x_3,\cdots, x_{n+1})
\biggr ].
\label{zasym}
\ee
We evaluate these integrals by means of the orthogonal polynomial
method \cite{PZJ97}.
The orthonormal polynomials corresponding to the exponential weight function
$ e^{-2x}$ are the Laguerre polynomials
\be
P_k(x) =\sqrt 2 L_k(2x).
\ee
By adding rows and columns the Vandermonde determinants can be expressed
in terms of these orthogonal polynomials. Keeping only the leading
order terms in $1/\mu$ in the exponentially suppressed term, we arrive
at the partition function
\be
{\cal Z}_{{\rm QCD}_3}^{(2+2n)}
 &=& \frac {e^{2n \mu}}{2^{n^2} \mu^{n^2 +2n} }\biggl [
2(n+1)\int_{0}^\infty d x_2 \cdots \int_{0}^\infty d x_{n+1}
e^{-2 x_2 - \cdots -2x_{n+1}} {\det}^2[P_k(x_2),
\cdots, P_k(x_{n+1})] \biggr .\nonumber \\
&-&  \frac {n(n+1)}\mu \int_{0}^\infty d x_1
\cdots \int_{0}^\infty d x_{n+1}
e^{-x_1-x_2-2 x_3 - \cdots -2x_{n+1}} \nonumber \\
&&\times \det[P_k(x_1),P_k(x_3), \cdots, P_k(x_{n+1})]
\det[P_k(x_2),P_k(x_3), \cdots, P_k(x_{n+1})]
\nonumber \\
&+&\biggl .  \frac {2\sqrt 2
n(n+1) e^{-2\mu} (-4\mu)^{n-1}}{\mu (n-1)!}
\int_{0}^\infty d x_1
\cdots \int_{0}^\infty d x_{n+1}
e^{-x_1-x_2-2 x_3 - \cdots -2x_{n+1}} \nonumber \\
&&\times \det[P_k(x_3), \cdots, P_k(x_{n+1})]
\det[P_k(x_2),P_k(x_3), \cdots, P_k(x_{n+1})] \biggr ] +\cdots \, ,
\label{zasym2}
\ee
where the ellipses denotes higher order contribution in $1/\mu$ to   
the exponentially suppressed terms.                                  
Because the polynomials are normalized to one, the first integral simply gives
$n!$. This can be easily seen by writing the determinants as a sum over
permutations and performing the integrals by orthogonality.
In the second and third
integral we have to treat the integration over $x_1$ and
$x_2$ separately. In the same way as for the
first integral, the remaining integrals just give $(n-1)!$.
To do the integrations over $x_1$ and $x_2$ we need the integral
\be
\int_0^\infty dx e^{-x} P_k(x) = (-1)^k \sqrt 2.
\ee
In the third integral, only the terms that contain $P_{n-1}(x_2)$ are
nonvanishing. The integrals over $x_1$ and $x_2$ have to be performed
separately, and, by orthogonality, the integrals over $x_3, \cdots, x_{n+1}$
give a factor $(n-1)!$.
We thus find
\be
{\cal Z}_{{\rm QCD}_3}^{(2+2n)}
&=& \frac {e^{2n \mu}}{2^{n^2} \mu^{n^2 +2n} }\biggl [
2(n+1)! -
\frac{(n+1)!}\mu \int_0^\infty dx_1\int_0^\infty dx_2
e^{-x_1-x_2} \sum_{k=0}^{n-1}P_k(x_1)P_k(x_2) \nonumber \\
&-& \biggl . \frac {2\sqrt 2
(n+1)! e^{-2\mu} (4\mu)^{n-1}}{(n-1)!\mu}
 \int_0^\infty dx_1\int_0^\infty dx_2
e^{-x_1-x_2} P_{n -1}(x_2) \biggr ]
\nonumber \\
&=&\frac {e^{2n\mu}}{2^{n^2} \mu^{n^2 +2n} }2(n+1)!
\left [1 -\frac{ n} \mu
+ \frac{2^{2n -1} (-1)^{n} \mu^{n-2}e^{-2\mu}  }{(n-1)!}\right ]+ \cdots \, .
\label{zasym3}
\ee
The valence quark mass dependence is thus given by
\be
\frac{\Sigma(\mu)}\Sigma &=& \lim_{n\to 0}\frac 1{2n} \partial_\mu
\log {\cal Z}_{{\rm QCD}_3}^{(2+2n)}
\nonumber \\ & = & 1-\frac 1\mu +\frac 1{2\mu^2} - \frac {e^{-2\mu}}{2\mu^2} ,
\ee
where we have calculated the replica limit of the exponentially suppressed
term by writing $1/(n-1)! = n/ n!$.  Notice that only the leading
order term in the expansion in $1/\mu$ has been included in the
exponentially suppressed term.

\noi
The exact result for the microscopic spectral density for one massless
flavor is given by
\be
\rho_s(u) = \frac 1\pi \left [ 1 - \frac {\sin^2 u}{u^2} \right ].
\ee
The corresponding valence quark mass dependence is equal to
\be
\frac{\Sigma(\mu)}\Sigma &=& \int_0^\infty\frac{2\mu\rho_s(u)} {u^2 +\mu^2} du
\nonumber \\           &=& 1 -\frac 1\mu +\frac 1{2\mu^2}(1-e^{-2\mu}),
\ee
in perfect agreement with the replica result. All higher order contributions
to the exponentially suppressed term cancel!

\noi
Let us now consider the case of two massless flavors.
We again expand the determinant in (\ref{detter}) and collect terms that
result in the same value for the integral as we did in the case of one
massless flavor. We find
\be
{\cal Z}_{{\rm QCD}_3}^{(4+2n)}
 &=& \frac 1{\mu^{4n}} \frac 1{(\prod_{k=1}^{n-1} k!)^2}
\int_{-1}^1 \!d \lambda_1 ...\int_{-1}^1 \!d \lambda_{n+2}
\biggl [{ n+2 \choose 2}
 e^{\mu(2\lambda_3 + \cdots +2\lambda_{n+2})}(\lambda_2-\lambda_1)^2
\Delta^2(\lambda_3, \cdots, \lambda_{n+2})
\biggr .
\nonumber \\&-&
2{ n+2 \choose 3} \
{ 3 \choose 2 } (\lambda_2 -\lambda_1)
(\lambda_3-\lambda_1)
e^{\mu(\lambda_2 + \lambda_3 +2\lambda_4 + \cdots +2\lambda_{n+2})}
\nonumber \\ &&\qquad \qquad \times\,\,
\Delta(\lambda_2, \lambda_4, \cdots, \lambda_{n+2})
\Delta(\lambda_3, \lambda_4, \cdots, \lambda_{n+2})
\nonumber \\&+&
{n+2 \choose 4 }{ 4 \choose 2 } (\lambda_2 -\lambda_1)
(\lambda_4-\lambda_3)
e^{\mu(\lambda_1+ \lambda_2 + \lambda_3 +\lambda_4+
2\lambda_5 + \cdots +2\lambda_{n+2})}
\nonumber \\ &&\qquad \qquad \times \,\,
\Delta(\lambda_1, \lambda_2, \lambda_5, \cdots, \lambda_{n+2})
\Delta(\lambda_3, \lambda_4,\lambda_5, \cdots, \lambda_{n+2})
\biggr ].
\label{zdelta}
\ee
The integrations that are not accompanied by an exponential should be
performed exactly, whereas the remaining integration variables are
rescaled as $\lambda_k = \pm 1 - x_k /\mu$ with sign determined by the upper
limit of the integration in (\ref{split}).
They are performed by expanding
the determinants as sums over permutations.
The integrations that are accompanied
by the exponential $e^{2x_k}$ are easily evaluated by orthogonality, and
give as result $n!$,\, $(n-1)!$ and $(n-2)!$, respectively, for the three
different types of terms in (\ref{zdelta}). Summarizing,
by keeping only the leading and first subleading
$1/\mu$ corrections in the
exponentially suppressed term,  we find
\be
{\cal Z}_{{\rm QCD}_3}^{(4+2n)}
 = \frac{e^{2n\mu}}{\mu^{4n+n^2} 2^{n^2}}
\biggl [ &&
{n+2 \choose 2} n!\int_{-1}^1 (\lambda_1 -\lambda_2)^2 d\lambda_1
d\lambda_2  \nonumber \\
-&&2{n+2 \choose 3} {3 \choose 2 }\frac{(n-1)!}\mu
\int_{-1}^1  d \lambda_1 \int_0^\infty\int_0^\infty
dx_2 dx_3 (1-\lambda_1- \frac {x_2}\mu)(1-\lambda_1- \frac {x_3}\mu)
\nonumber \\
&&\qquad   \times \,
\sum_{k=0}^{n-1} P_k(x_2)P_k(x_3)e^{-x_2-x_3}\nonumber \\
+&& {n+2 \choose 4} {4 \choose 2} (n-2)!\frac 1{\mu^4}
\int_0^\infty \cdots \int_0^\infty  dx_1 \cdots dx_4 (x_2-x_1)(x_4-x_3)
\nonumber \\
&&\qquad\times \, \sum_{k<l}P_k(x_1)P_l(x_2)[P_k(x_3)P_l(x_4)
- P_l(x_3)P_k(x_4)] e^{-x_1-x_2-x_3-x_4}  \nonumber \\
+&& 4{n+2 \choose 3} {3 \choose 2 }{\mu^{n-2}e^{-2\mu}
(-1)^{n-1}2^{2n-1} }
\int_{-1}^1  d \lambda_1 \int_0^\infty\int_0^\infty
dx_2 dx_3 \nn\\
&&\times (-1-\lambda_1-\frac{x_2}\mu)(1-\lambda_1-\frac{x_3}\mu)
e^{-x_2-x_3} \left ( \frac{P_{n-1}(x_3)}{{\sqrt 2}} \right . \nn \\
+&&\left.\frac{n-1}2\frac {x_2}\mu \frac{P_{n-1}(x_3)}{{\sqrt 2}}
-\sum_{k=0}^{n-2} \frac{2k+1}{4\mu} \frac{P_{n-1}(x_3)}{{\sqrt 2}}
-\frac{n-1}{4\mu} \frac{P_{n-2}(x_3)}{{\sqrt 2}}\right)
\biggr ] +\cdots . \nonumber \\                                             
\ee
In the last terms we have included a factor 2 to account for
expanding $\lambda_2$ and
$\lambda_3$ around $+1$ and $-1$ and vice versa. The last three terms represent
the $1/\mu$ corrections to the Vandermonde determinant.  They have
been simplified by means of recursion relations for the Laguerre polynomials.
The ellipses denote higher order corrections to the exponentially     
suppressed terms.                                                     
To calculate the integrals we need the following result:
\be
\int_0^\infty dx e^{-x} x P_k(x) = (-1)^k (2k+1)\sqrt 2.
\ee
The partition function is thus given by
\be
{\cal Z}_{{\rm QCD}_3}^{(4+2n)}
 = \frac{e^{2n \mu}}{\mu^{4n+n^2} 2^{n^2}}\biggl [ &&
{ n+2 \choose 2} \frac {8n!}3\nn\\
-&&2{ n+2 \choose 3} { 3 \choose 2 } (n-1)!
\biggl [  \frac {16n}{3\mu} + \frac 8{\mu^2} \sum_{k=0}^{n-1}(2k+1)
+\frac 4{\mu^3} \sum_{k=0}^{n-1}(2k+1)^2
\biggr ] \nonumber \\
+&& {n+2 \choose 4 }{ 4 \choose 2 }\frac{8(n-2)!}{\mu^4}
\sum_{k<l}((-1)^k 2k - (-1)^l 2l)^2 \nn\\
-&& e^{-2\mu}{ n+2 \choose 3} { 3 \choose 2 }\frac{8}{3\mu^{2-n}}
\left ( 1 - \frac {(n-1)(2+n/4)}\mu \right )
\biggr ] +\cdots .
\ee
The sums over $k$ and $l$ are elementary,
\be
\sum_{k=0}^{n-1} (2k+1) &= & n^2 ,\\
\sum_{k=0}^{n-1} k^2 &= &\frac n6 (n-1)(2n-1), \\
\sum_{k=0}^{n-1} (2k+1)(-1)^k &= & n (-1)^{n+1},\\
\sum_{k=0}^{n-1} (2k+1)^2 &= &\frac 43 n (n^2 - \frac 14),\\
\sum_{k,l=0}^{n-1} [(2k+1)(-1)^k- (2k+1)(-1)^l ]^2&= & \frac 83 n^2 (n^2-1).
\ee
We thus find the partition function
\be
{\cal Z}_{{\rm QCD}_3}^{(4+2n)}
 &=& \frac{4(n+2)! e^{2n \mu}}{2^{n^2}3\mu^{4n+n^2} }
\biggl [1 - \frac {4n}\mu +\frac {6n^2}{\mu^2}
-\frac {4}{\mu^3} n (n^2 -\frac 14)
+ \frac {n^2}{\mu^4} (n^2-1) \nn\\
&&-\ \frac{ne^{-2\mu}}{\mu^2}\left ( 1 - \frac {(n-1)(2+n/4)}\mu \right )
\biggr ] +\cdots \ .
\ee
In the replica limit this reduces to
\be
{\cal Z}_{{\rm QCD}_3}^{(4+2n)}
 = \frac{4e^{2n \mu}}{3\mu^{4n} }
\biggl [ 1 - \frac {4n}\mu +\frac {n}{\mu^3} -ne^{-2\mu}
\left (\frac 1{\mu^2} + \frac 2{\mu^3} \right )
\biggr ] +\cdots ,
\ee
which leads to the following $\mu$-dependence of the resolvent
\be
\frac{\Sigma(\mu)}\Sigma = 1 - \frac 2\mu+ \frac 2{\mu^2} -\frac 3{2\mu^4}
+ e^{-2\mu}\left (\frac 1{\mu^2} + \frac 3{\mu^3} + \frac{3}{2\mu^4}
\right )
+\cdots \quad .
\label{cond2nf}
\ee
We have not calculated the terms of order $e^{-2\mu}/\mu^4$ that appear
in the expansion of the partition function. By comparison with the
exact result given by
\be
\frac{\Sigma(\mu)}\Sigma &=&\frac \mu2 \left [I_{5/2}(\mu)K_{5/2}(\mu)+
I_{3/2}(\mu)K_{3/2}(\mu)+I_{5/2}(\mu)K_{1/2}(\mu)+ I_{7/2}(\mu)
K_{3/2}(\mu)\right]
\nonumber \\
&=& 1 - \frac 2\mu +  \frac 2{\mu^2} - \frac 3{2\mu^4}
+e^{-2\mu} \left [ \frac 1{\mu^2} + \frac 3{\mu^3} + \frac 3{2\mu^4} \right ],
\ee
it turns out that such terms vanish in the replica limit (see also
eq. (\ref{QCD3resolvent})). To complete the
calculation we will have to show that these terms, as well as higher order
exponentially suppressed terms, vanish in the replica limit.
The termination of the asymptotic series we have observed in this calculation
is actually a more general phenomenon known as Duistermaat-Heckman
localization \cite{DM,Z1} (see \cite{Szabodm} for a pedagogical discussion
of this topic).

\newpage
\section{Conclusions}
\vspace*{0.3cm}

\noi
In the domain where the Compton wavelength of the Goldstone modes
is much larger than the size
of the box, the QCD$_3$ finite volume partition function for
$2N_f$ flavors is given by a $U(2N_f)/ U(N_f)\times U(N_f)$
unitary matrix integral. The main objective  of this article
has been to relate the spectrum of the QCD$_3$ Dirac
operator to this partition function, and extensions of it with more flavors.
An alternative starting point would have been the low-energy limit
of the graded (or supersymmetric) generating function for the Dirac spectrum.
This partition function is obtained by complementing the
QCD partition function with both fermionic and bosonic ``ghost quarks". It is
invariant under an internal supersymmetry.
In this way correlation functions of the QCD Dirac eigenvalues can be obtained
in a mathematically rigorous way. However, we have here
been interested in the question whether the usual low-energy effective
partition function, without bosonic ghost quarks,
already contains sufficient information to completely
constrain the low-energy Dirac spectrum.

\noi
This question has been answered affirmatively earlier within the context of
Random Matrix Theory. The spectral $k$-point correlation functions are
given by the ratio of the usual partition function and a
partition function with $2k$ additional flavors. These additional
flavors can be related to the fermionic and bosonic ghost quarks
of the supersymmetric generating function. {\em A priori} there is
no reason why the spectral correlation functions of eigenvalues in Random
Matrix Theory should coincide with those of the Dirac operator in the
low-energy limit of QCD. Can we reach the same
conclusion without relying on Random Matrix Theory?

\noi
We have studied this question in three different ways. First, we have
found exact relations between the QCD$_3$ partition function and
the QCD$_4$ partition function analytically continued to half-integer
topological charge. We have next added $n$ replica quarks to the two theories
and taken the replica limit. This has allowed us to derive explicit relations
between the microscopic spectral density of the Dirac
operator in QCD$_4$ and
the corresponding microscopic spectral density of the Dirac operator
in QCD$_3$ . These relations suggest the
existence of not yet well-understood connections between the physics
of QCD$_3$ and QCD$_4$, and are thus of independent interest.
Second, the finite volume partition function of QCD$_3$
satisfies Virasoro constraints. They can
be solved recursively, providing us with the small mass expansion of the
finite volume partition function, and a series of spectral sum rules.
Third, we have directly analyzed the replica limit
of the Unitary Matrix integral corresponding to the finite-volume
partition functions of QCD$_3$. In this way we have been able to derive the
first few terms of the large mass expansion. Because the series truncate,
this expansion turns out to be exact.

\noi
A more general question we have addressed
is the applicability of the replica method to                          
the calculation of quenched
averages. It was suggested recently that exact non-perturbative results
can be obtained using an appropriate scheme of replica symmetry breaking.
The general relations we have derived between the microscopic spectral
densities
of QCD$_3$ and QCD$_4$, though, follow from a completely straightforward
replica limit.
As an example,
either by using previously obtained results for the QCD$_4$ partition function
or by the direct replica calculation of this paper we have reproduced
exact results from the large mass expansion of the partition function.
Possible caveats of the replica limit do not show up in our calculation.   
We do not know the reasons behind this observation, and               
it would be interesting to obtain a better understanding of the conditions 
for the applicability of the replica method.                              

\noi
Generally, it has been expected that the replica method will not give
exact results for spectral correlation functions but, at best,
the coefficients of an asymptotic expansion.
However, there are important cases where the series expansion turns out
to {\em truncate}.
We have witnessed a dramatic example of this phenomenon in this article.
However, there are still subtleties. The
finite volume QCD$_4$ partition function is well-defined for all integer
values $\nu$ of the topological charge.
What is its analytical continuation in the complex $\nu$-plane?
A naive analytical continuations fails because of the presence
of a $\sin \pi \nu$-term. The correct analytical continuation is obtained
by changing the integration contour such that translational invariance
is restored.
We have not calculated for general $N_f$ the order at which the
large mass expansion terminates. We believe that it is important to get
a better understanding of this question and its relation with
Duistermaat-Heckman localization, but this is beyond the scope of this work.

\noi
The replica method also yields the correct power-series expansion for small
masses. As in
QCD$_4$ the small mass expansion has de Wit--'t Hooft poles that limit
the number of terms that can be obtained this way, but up to the order
the replica method can be used all results agree exactly
with earlier results from both Random Matrix Theory and the supersymmetric
formulation. These three very different methods are thus obviously confirming
each other.

\noi
In conclusion, we have shown the existence of powerful relations between
the QCD$_3$ and the QCD$_4$ partition functions. The replica limit of these
results provides us with exact non-perturbative relations between the spectral
correlation functions of the Dirac operator in the two theories.
It would be most interesting to derive similar relations
from the supersymmetric formulation as well.

\noi\noindent
{\bf Acknowledgments}\\
This work was partially supported by the US DOE grant
DE-FG-88ER40388.
The work of G.A. and P.H.D. was supported in part by EU TMR grant no.
ERBFMRXCT97-0122. D.D. is supported by FAPESP (Brazilian Agency).
Kim Splittorff and Dominique Toublan are thanked for useful discussions. 
Two of us (P.H.D. and J.J.M.V.)
acknowledge useful discussions at the Institute of Nuclear
Theory  of the University of Washington
at Seattle in the initial stages of this work.

\vspace{0.6cm}
\newpage
\appendix
\setcounter{equation}{0}

\section{Relations between QCD$_3$ and QCD$_4$ determinants}
\label{3-4dets}
\vspace*{0.3cm}

\noi
In this Appendix we will prove the two eqs. (\ref{detrel+}) and
(\ref{detrel-}) from two different lemmas, which then completes the
proof of Theorem III eq. (\ref{factor3}).

\noi\noindent
{\sc Lemma 1}: {\it Let the matrices $\matD (\{\mu_i\},\zeta,\om)$,
$\matC (\{\mu_i\},\zeta,\om)$ and $\matS (\{\mu_i\})$ be defined as in
eqs. (\ref{matDdef}) and (\ref{rhseven}), respectively.
Then the following relation holds:}
\beq
(\om+\zeta)\det {\matD}(\{\mu_i\},\zeta,\om) +
(\om-\zeta)\det {\matD}(\{\mu_i\},-\zeta,\om)
= (-1)^{N_f+1}
2^{N_f+1}\det\matC(\{\mu_i\},\zeta,\om)\det\matS(\{\mu_i\}).
\label{La1}
\eeq

\noi\noindent
{\sc Proof}: We will show that the left hand side factorizes into the
right hand side by adding and subtracting rows and columns and
expanding the determinants appropriately.

\noi
Due to the block structure in the definition of the matrix ${\matD}$ in
eq. (\ref{matDdef}) we can add the upper blocks $\matC$ and $\matS$
to the lower ones and obtain after taking out factors of two,
\beq
\det {\matD}(\{\mu_i\},\zeta,\om)\ =\
2^{N_f}\det\left(
\begin{array}{ll}
\matC(\{\mu_i\})  & -\matS(\{\mu_i\})  \\
\matC(\{\mu_i\})  &\ \ {\bf 0}\\
\zeta^{j-1}\cosh^{(j-1)} (\zeta) &-\zeta^{j-1}\sinh^{(j-1)} (\zeta) \\
\om^{j-1}\cosh^{(j-1)} (\om) &-\om^{j-1}\sinh^{(j-1)} (\om)
\end{array}\right) \ .
\label{D0block1}
\eeq
 When evaluating the determinant
${\matD}(\{\mu_i\},-\zeta,\om)$ we can use that in the row depending on
$\zeta$ the functions $\zeta^{j-1}\cosh^{(j-1)}(\zeta)$
are even under $\zeta\to-\zeta$ whereas the functions
$\zeta^{j-1}\sinh^{(j-1)} (\zeta)$ are odd. In order to add the two
determinants in eq. (\ref{La1}) we multiply the factors $(\om\pm\zeta)$
into the last column of the respective determinants.
Next, we use that determinants differing by only one single
column or row can be added. We use this to split the left hand side of
eq. (\ref{La1}) in the following way, where we explicitly display the
last column,
\beqn
(\om+\zeta)\det {\matD}(\{\mu_i\},\zeta,\om) &&+\ \ \
  (\om-\zeta)\det {\matD}(\{\mu_i\},-\zeta,\om)  \ =\     \label{start}\\
\ =\ 2^{N_f} &&\left[
\det\left(
\begin{array}{lll}
\matC(\{\mu_i\})  & -\matS(\{\mu_i\}) & -\om\mu_i^{N_f}\sinh^{(N_f)}(\mu_i) \\
\matC(\{\mu_i\})  &\ \ {\bf 0}        & \ \ 0\\
\zeta^{j-1}\cosh^{(j-1)} (\zeta)      &-\zeta^{j-1}\sinh^{(j-1)} (\zeta)
                                      &-\zeta^{N_f+1} \sinh^{(N_f)}(\zeta)\\
\om^{j-1}\cosh^{(j-1)} (\om)          &-\om^{j-1}\sinh^{(j-1)} (\om)
                                      &-\om^{N_f+1} \sinh^{(N_f)}(\om)
\end{array}\right)
\right. \nn\\
&&+
\det\left(
\begin{array}{lll}
\matC(\{\mu_i\})  & -\matS(\{\mu_i\})& -\zeta\mu_i^{N_f}\sinh^{(N_f)}(\mu_i)\\
\matC(\{\mu_i\})  &\ \ {\bf 0}       & \ \ 0\\
\zeta^{j-1}\cosh^{(j-1)} (\zeta)      &-\zeta^{j-1}\sinh^{(j-1)} (\zeta)
                                      &-\om\zeta^{N_f} \sinh^{(N_f)}(\zeta)\\
\om^{j-1}\cosh^{(j-1)} (\om)          &-\om^{j-1}\sinh^{(j-1)} (\om)
                                      &-\zeta\om^{N_f} \sinh^{(N_f)}(\om)
\end{array}\right)
\nn\\
&&+
\det\left(
\begin{array}{lll}
\matC(\{\mu_i\})  & -\matS(\{\mu_i\}) & -\om\mu_i^{N_f}\sinh^{(N_f)}(\mu_i) \\
\matC(\{\mu_i\})  &\ \ {\bf 0}        & \ \ 0\\
\zeta^{j-1}\cosh^{(j-1)} (\zeta)      &\zeta^{j-1}\sinh^{(j-1)} (\zeta)
                                      &-\zeta^{N_f+1} \sinh^{(N_f)}(\zeta)\\
\om^{j-1}\cosh^{(j-1)} (\om)          &-\om^{j-1}\sinh^{(j-1)} (\om)
                                      &-\om^{N_f+1} \sinh^{(N_f)}(\om)
\end{array}\!\right)
\nn\\
&&+ \left.
\det\left(
\begin{array}{lll}
\matC(\{\mu_i\})  & -\matS(\{\mu_i\})&\zeta\mu_i^{N_f}\sinh^{(N_f)}(\mu_i)\\
\matC(\{\mu_i\})  &\ \ {\bf 0}       &\ \ 0\\
\zeta^{j-1}\cosh^{(j-1)} (\zeta)     &\zeta^{j-1}\sinh^{(j-1)} (\zeta)
                                     &\om\zeta^{N_f} \sinh^{(N_f)}(\zeta)\\
\om^{j-1}\cosh^{(j-1)} (\om)         &-\om^{j-1}\sinh^{(j-1)} (\om)
                                     &\zeta\om^{N_f} \sinh^{(N_f)}(\om)
\end{array}\right)
\!\right]\!. \nn
\eeqn
In the next step we can add the first and the third determinant on the
right hand side, since they only differ by the last but one row.
And we can also add the second and forth determinant for the same
reason, after multiplying in the fourth determinant the last column and
last but one row with $(-1)$. We find
\beqn
(\om+\zeta)\det {\matD}(\{\mu_i\},\zeta,\om) &+&
  (\om-\zeta)\det {\matD}(\{\mu_i\},-\zeta,\om)  \ =\   \label{detstep1}   \\
\ =\  2^{N_f+1}&&\left[
\det\left(
\begin{array}{lll}
\matC(\{\mu_i\})  & -\matS(\{\mu_i\}) & \ \ 0\\
\matC(\{\mu_i\})  &\ \ {\bf 0}        & \ \ 0\\
\zeta^{j-1}\cosh^{(j-1)} (\zeta)      & \ \ 0
                                      &-\zeta^{N_f+1} \sinh^{(N_f)}(\zeta)\\
\om^{j-1}\cosh^{(j-1)} (\om)          &-\om^{j-1}\sinh^{(j-1)} (\om)
                                      &-\om^{N_f+1} \sinh^{(N_f)}(\om)
\end{array}\!\right)\right.
\nn\\
+&&
\det\left(
\begin{array}{lll}
\matC(\{\mu_i\})  & -\matS(\{\mu_i\}) & -\om\mu_i^{N_f}\sinh^{(N_f)}(\mu_i) \\
\matC(\{\mu_i\})  &\ \ {\bf 0}        & \ \ 0\\
\zeta^{j-1}\cosh^{(j-1)} (\zeta)      & \ \ 0
                                      & \ \ 0\\
\om^{j-1}\cosh^{(j-1)} (\om)          &-\om^{j-1}\sinh^{(j-1)} (\om)
                                      & \ \ 0
\end{array}\right)\nn\\
+&&\left.
\det\left(
\begin{array}{lll}
\matC(\{\mu_i\})  & -\matS(\{\mu_i\})& -\zeta\mu_i^{N_f}\sinh^{(N_f)}(\mu_i)\\
\matC(\{\mu_i\})  &\ \ {\bf 0}       & \ \ 0\\
0                                    &-\zeta^{j-1}\sinh^{(j-1)} (\zeta)
                                     & \ \ 0\\
\om^{j-1}\cosh^{(j-1)} (\om)         &-\om^{j-1}\sinh^{(j-1)} (\om)
                                     &-\zeta\om^{N_f} \sinh^{(N_f)}(\om)
\end{array}\right)
\!\right], \nn
\eeqn
where we have again split the
determinant, resulting from the first addition, into two parts with
respect to the last column,
and taken out a factor of 2. The first determinant in the last
equation is already part of the desired result. In the last column we
can write $\sinh^{(N_f)}=\cosh^{(N_f+1)}$. If we then permute the last
column $N_f$ times to the left the first $N_f+2$ columns only contain
derivatives of $\cosh$. We Laplace-expand the resulting determinant
into products of $(N_f+2)\times(N_f+2)$ blocks and
$N_f\times N_f$ blocks. After taking out all signs we obtain
\beq
(-1)^{N_f+1} 2^{N_f+1}
\det\left(
\begin{array}{ll}
\matC(\{\mu_i\})                & 0\\
\zeta^{j-1}\cosh^{(j-1)}(\zeta) &\zeta^{N_f+1} \cosh^{(N_f+1)}(\zeta)\\
  \om^{j-1}\cosh^{(j-1)}( \om)  &\om^{N_f+1} \cosh^{(N_f+1)}(\om)
\end{array}
\right)
\det\matS(\{\mu_i\})\ .
\label{detpiece1}
\eeq
What remains to be shown is that the two
remaining determinants in eq. (\ref{detstep1}) give rise to the same
term as in eq. (\ref{detpiece1}), except that the last column vector
is reading $(\mu_i^{N_f+1}\cosh^{(N_f+1)}(\mu_i),0,0)$. In
that way they will add up to the right hand side of
Lemma 1, eq. (\ref{La1}).

\noi
To proceed, we find that in the third determinant of eq. (\ref{detstep1}), the
last term in the last column can be put equal to zero. To
see this we expand the determinant with respect to the last column and
note that the determinant multiplying $-\zeta\om^{N_f}
\sinh^{(N_f)}(\om)$ vanishes, which can be easily seen by expanding
it into products of $(N_f+1)\times(N_f+1)$ blocks and $N_f\times N_f$ blocks.
Next, in both the second and third determinant
in  eq. (\ref{detstep1}) we take the common factors $\zeta$ and
$\om$ out of the last column, respectively, and multiply them into the
last but one row. The two determinants are then equal up to precisely
the last but one row and can thus be added. We arrive at
\beq
2^{N_f+1} \det\left(
\begin{array}{lll}
\matC(\{\mu_i\})  & -\matS(\{\mu_i\}) & -\mu_i^{N_f}\sinh^{(N_f)}(\mu_i) \\
\matC(\{\mu_i\})  &\ \ {\bf 0}        & \ \ 0\\
\om\zeta^{j-1}\cosh^{(j-1)} (\zeta)   & -\zeta^{j}\sinh^{(j-1)} (\zeta)
                                      & \ \ 0\\
\om^{j-1}\cosh^{(j-1)} (\om) &-\om^{j-1}\sinh^{(j-1)} (\om)
                                      & \ \ 0
\end{array}\right).
\eeq
It is straightforward to see that, when expanding with respect to the
last column, and then into products of $(N_f+1)\times(N_f+1)$ blocks of $\matC$
and $N_f\times N_f$
of $\matS$, the $\om$ multiplying the left side of the last but
one row can be multiplied into the right side of the last row
\beq
2^{N_f+1} \det\left(
\begin{array}{lll}
\matC(\{\mu_i\})  & -\matS(\{\mu_i\}) & -\mu_i^{N_f}\sinh^{(N_f)}(\mu_i) \\
\matC(\{\mu_i\})  &\ \ {\bf 0}        & \ \ 0\\
\zeta^{j-1}\cosh^{(j-1)} (\zeta)      & -\zeta^{j}\sinh^{(j-1)} (\zeta)
                                       & \ \ 0\\
\om^{j-1}\cosh^{(j-1)} (\om)            &-\om^{j}\sinh^{(j-1)} (\om)
                                       & \ \ 0
\end{array}\right).
\label{detstep2}
\eeq
As we have said before we claim that this is equal to the following product:
\beq
(-1)^{N_f+1} 2^{N_f+1}
\det\left(
\begin{array}{ll}
\matC(\{\mu_i\})                & \mu_i^{N_f+1} \cosh^{(N_f+1)}(\mu_i)\\
\zeta^{j-1}\cosh^{(j-1)}(\zeta) &\ \ 0 \\
  \om^{j-1}\cosh^{(j-1)}( \om)  &\ \ 0
\end{array}
\right)
\det\matS(\{\mu_i\})\ ,
\label{detpiece2}
\eeq
which will add up to eq. (\ref{detpiece1}) in order to give the
desired right hand side of
eq. (\ref{La1}). To see this we still have to do two more Laplace expansions.
First of all we eliminate in  eq. (\ref{detstep2}) the upper block
$\matC$ by subtracting the lower one. Then, we permute the first
column $N_f$ times to the right to obtain
\beq
-2^{N_f+1} \det\left(
\begin{array}{llll}
\ \ {\bf 0}                  & \ \ 0      &\matS(\{\mu_i\})
                                          &\mu_i^{N_f}\sinh^{(N_f)}(\mu_i)\\
\mu_i^{j}\cosh^{(j)}(\mu_i)  &\cosh(\mu_i)&{\bf 0}
                                          &0\\
\zeta^{j}\cosh^{(j)} (\zeta) &\cosh(\zeta)&\zeta^{j}\sinh^{(j-1)} (\zeta)
                                          &0\\
\om^{j}\cosh^{(j)} (\om)     &\cosh(\om)  &\om^{j}\sinh^{(j-1)} (\om)
                                          &0
\end{array}\right),
\label{detstep3}
\eeq
after taking out all signs.
Our strategy will now be as follows. We will expand this determinant
into products of  $N_f\times N_f$ blocks
and $(N_f+2)\times(N_f+2)$ blocks. The first blocks, after taking
out common factors and rewriting $\cosh^{(j)}=\sinh^{(j-1)}$ will
give rise to the desired prefactor $\det\matS((\{\mu_i\})$. In doing so
we obtain for eq. (\ref{detstep3})
\beqn
&&(-1)^{N_f+1}2^{N_f+1}
\left\{\det\left(\mu_i^j\cosh^{(j)}(\mu_i)\right)
\det\left(
\begin{array}{lll}
\ \ 0       &\matS(\{\mu_i\})             &\mu_i^{N_f}\sinh^{(N_f)}(\mu_i)\\
\cosh(\zeta)&\zeta^{j}\sinh^{(j-1)}(\zeta)& \ \ 0\\
\cosh(\om)  &\om^{j}\sinh^{(j-1)} (\om)   & \ \ 0
\end{array}\right) \right.
\label{detstep4}\\
&&\ \ \ \ -\sum_{k=1}^{N_f}(-1)^k \left[
\det\left(
\begin{array}{l}
\mu_i^j\cosh^{(j)}(\mu_i)\\
\ldots \mbox{no}\ k\ldots\\
\om^{j}\cosh^{(j)}(\om)
\end{array}\right)
\det\left(
\begin{array}{lll}
\cosh(\mu_k)&\ \ {\bf 0}                  & \ \ 0\\
\ \ 0        &\matS(\{\mu_i\})            &\mu_i^{N_f}\sinh^{(N_f)}(\mu_i)\\
\cosh(\zeta)&\zeta^{j}\sinh^{(j-1)}(\zeta)& \ \ 0\\
\end{array}\right)
\right.
\nn\\
&&\ \ \ \ \ \ \ \ \ \ \ \ \ \ \ \ -\ \left.\left.
\det\left(
\begin{array}{l}
\mu_i^j\cosh^{(j)}(\mu_i)\\
\ldots \mbox{no}\ k\ldots\\
\zeta^{j}\cosh^{(j)}(\zeta)
\end{array}\right)
\det\left(
\begin{array}{lll}
\cosh(\mu_k)&\ \ {\bf 0}                  & \ \ 0\\
\ \ 0       &\matS(\{\mu_i\})             &\mu_i^{N_f}\sinh^{(N_f)}(\mu_i)\\
\cosh(\om)  &\om^{j}\sinh^{(j-1)}(\om)    & \ \ 0\\
\end{array}\right)
\right]\right\}. \nn
\eeqn
The first product gives already part of eq. (\ref{detpiece2}) whereas the
remaining contributions originate from the other terms. To see
that we take out $\prod_i \mu_i$ out of the first determinant
and multiply the $i$th row of the second
determinant with $\mu_i$ (for $i = 1, \cdots, N_f)$.
Then, we write $\cosh^{(j)}=\sinh^{(j-1)}$ in the first and
$\sinh^{(j-1)}=\cosh^{(j)}$ in the second factor to obtain
\beq
(-)^{N_f+1}2^{N_f+1}\det\matS(\{\mu_i\})\
\det\left(
\begin{array}{lll}
\ \ 0       &\mu_i^{j}\cosh^{(j)}(\mu_i)&\mu_i^{N_f+1}\cosh^{(N_f+1)}(\mu_i)\\
\cosh(\zeta)&\zeta^{j}\cosh^{(j)}(\zeta)& \ \ 0\\
\cosh(\om)  &\om^{j}\cosh^{(j)} (\om)   & \ \ 0
\end{array}\right) .
\label{detpiece2a}
\eeq
We are still left with the sum over $k$ in eq. (\ref{detstep4}). To
simplify it further we take out common factors $\mu_{i\neq k}$
from the first determinant as well as rewrite
$\cosh^{(j)}=\sinh^{(j-1)}$.
We then expand the second determinant into products of $2\times2$ blocks and
$N_f\times N_f$ blocks, where the $2\times2$
blocks consist of the first and last column
\beqn
&&2^{N_f+1}\sum_{k,l=1}^{N_f}(-1)^{k+l-1}
\left[
\det\left(
\begin{array}{l}
\mu_i^j\sinh^{(j-1)}(\mu_i)\\
\ldots \mbox{no}\ k\ldots\\
\om^{j}\sinh^{(j-1)}(\om)
\end{array}\right)
\cosh(\mu_k)\ \mu_l^{N_f}\!\sinh^{(N_f)}(\mu_l)
\det\left(
\begin{array}{l}
\matS(\{\mu_{i\neq l}\})\\
\zeta^{j}\sinh^{(j-1)}(\zeta)
\end{array}\right)
\right.\nn\\
&&\left.
\ \ \ \ \ \ \ \ \ \ \ \ \ \ \ \ \ \ \ \ \ \ \ \ -
\det\left(
\begin{array}{l}
\mu_i^j\sinh^{(j-1)}(\mu_i)\\
\ldots \mbox{no}\ k\ldots\\
\zeta^{j}\sinh^{(j-1)}(\zeta)
\end{array}\right)
\cosh(\mu_k)\ \mu_l^{N_f}\!\sinh^{(N_f)}(\mu_l)
\det\left(
\begin{array}{l}
\matS(\{\mu_{i\neq l}\})\\
\om^{j}\sinh^{(j-1)}(\om)
\end{array}\right)
\right]\nn\\
&&=\ 2^{N_f+1}\mbox{\Huge[}
\sum_{k,l=1}^{N_f}(-1)^{k+l-1}
\zeta\om\!\prod_{j\neq k,l}\mu_j\
\cosh(\mu_k)\ \mu_l^{N_f+1}\cosh^{(N_f+1)}(\mu_l)
\nn\\
&&\ \ \ \ \ \ \ \ \ \ \times\left(
\det\matS(\{\mu_{i\neq k}\},\om) \det\matS(\{\mu_{i\neq l}\},\zeta)
\ -\
\det\matS(\{\mu_{i\neq k}\},\zeta) \det\matS(\{\mu_{i\neq l}\},\om)
\right)\mbox{\Huge]}\nn\\
&&=\ 2^{N_f+1}\left[
\sum_{k>l}^{N_f}(-1)^{k+l-1}
\zeta\om\!\prod_{j\neq k,l}\mu_j\
\cosh(\mu_k)\ \mu_l^{N_f+1}\!\cosh^{(N_f+1)}(\mu_l)
\det\matS(\{\mu_{i}\}) \det\matS(\{\mu_{i\neq l,k}\},\zeta,\om)\right.\nn\\
&&\ \ \ \ \ \ \ \ \ \ - \left.
\sum_{k<l}^{N_f}(-1)^{k+l-1}
\zeta\om\!\prod_{j\neq k,l}\mu_j\
\cosh(\mu_k)\ \mu_l^{N_f+1}\!\cosh^{(N_f+1)}(\mu_l)
\det\matS(\{\mu_{i}\}) \det\matS(\{\mu_{i\neq k,l}\},\zeta,\om)\right]\nn\\
&&=\ 2^{N_f+1}\det\matS(\{\mu_{i}\}) \nn\\
&&\ \ \ \ \times\sum_{k>l}^{N_f}(-1)^{k+l-1}
\zeta\om\!\prod_{j\neq k,l}\mu_j
\det\left(
\begin{array}{ll}
\cosh(\mu_k)&\mu_k^{N_f+1}\!\cosh^{(N_f+1)}(\mu_k)\\
\cosh(\mu_l)&\mu_l^{N_f+1}\!\cosh^{(N_f+1)}(\mu_l)
\end{array}\right)
\det\matS(\{\mu_{i\neq k,l}\},\zeta,\om)
\nn\\
&&=\ (-1)^{N_f+1}2^{N_f+1}
\det\matS(\{\mu_i\})\
\det\left(
\begin{array}{lll}
\cosh(\mu_i)&\mu_i^{j}\cosh^{(j)}(\mu_i)
                                       &\mu_i^{N_f+1}\cosh^{(N_f+1)}(\mu_i)\\
\ \ 0       &\zeta^{j}\cosh^{(j)}(\zeta)& \ \ 0\\
\ \ 0       &\om^{j}\cosh^{(j)} (\om)     & \ \ 0
\end{array}\right) .
\label{detpiece2b}
\eeqn
In the first step we have only taken out common factors $\zeta$ and
$\om$ in order to be able to use the short notation $\matS$.
In the second step we have used a property for antisymmetric products
of determinants. A proof of the relations used can be found for
example in Lemma 5 of \cite{AD2}. Moreover, we have to distinguish
the cases $k>l$ and $k<l$, whereas the terms with $k=l$ drop out.
In step three we have rewritten everything in terms of
determinants. This is just to see that the double sum gives the
Laplace expansion of the matrix in the last line. Here,
we have again used that
$\sinh^{(j-1)}=\cosh^{(j)}$ and multiplied the factor
$\zeta\om\prod_{j\neq k,l}\mu_j$ into the last determinant
$\det\matS(\{\mu_{i\neq k,l}\},\zeta,\om)$. To summarize our efforts,
the last line of eq. (\ref{detpiece2b}) together with
eq. (\ref{detpiece2a}) sums up to eq. (\ref{detpiece2}), as we wished
to show. So we see that starting with the left hand side of
eq. (\ref{start}) we arrive at eqs. (\ref{detpiece1}) and
(\ref{detpiece2}), which add up to the right hand side of the lemma
eq. (\ref{La1}).

\vspace{0.6cm}
\noi\noindent
{\sc Lemma 2}:
{\it With the same definitions as in Lemma 1 the following relation holds:}
\beq
(\om+\zeta)\det {\matD}(\{\mu_i\},\zeta,\om) -
(\om-\zeta)\det {\matD}(\{\mu_i\},-\zeta,\om)
= (-1)^{N_f}
2^{N_f+1}\det\matS(\{\mu_i\},\zeta,\om)\det\matC(\{\mu_i\}).
\label{La2}
\eeq

\noi\noindent
{\sc Proof}: The proof will go very much along the same lines as in
Lemma 1. However, we will to arrive at a determinant of $\matS$ of
size $(N_f+2)$ instead of size $N_f$, which requires  a different addition
and subtraction scheme  of rows and columns than in the proof of Lemma 1.

\noi
We start again from the definition of the matrix ${\matD}$ in
eq. (\ref{matDdef}) and this time subtract the upper blocks $\matC$ and $\matS$
from the lower ones,
\beq
\det {\matD}(\{\mu_i\},\zeta,\om)\ =\
2^{N_f}\det\left(
\begin{array}{ll}
{\bf 0}           & -\matS(\{\mu_i\})  \\
\matC(\{\mu_i\})  &\ \ \matS(\{\mu_i\})  \\
\zeta^{j-1}\cosh^{(j-1)} (\zeta) &-\zeta^{j-1}\sinh^{(j-1)} (\zeta) \\
\om^{j-1}\cosh^{(j-1)} (\om) &-\om^{j-1}\sinh^{(j-1)} (\om)
\end{array}\right) \ .
\label{D0block2}
\eeq
If we multiply the factors $(\om\pm\zeta)$ into the $(N_f+1)$-st column of
the determinants and split them up the same way as before we obtain
\beqn
(\om+\zeta)\det {\matD}(\{\mu_i\},\zeta,\om) &&-\ \ \
  (\om-\zeta)\det {\matD}(\{\mu_i\},-\zeta,\om)  \ =\     \label{start2}\\
\ =\ 2^{N_f} &&\left[
\det\left(
\begin{array}{lll}
    {\bf 0}       & 0                                 & -\matS(\{\mu_i\})  \\
\matC(\{\mu_i\})  & \om\mu_i^{N_f}\cosh^{(N_f)}(\mu_i)&\ \ \matS(\{\mu_i\})\\
\zeta^{j-1}\cosh^{(j-1)} (\zeta)      &\zeta^{N_f+1} \cosh^{(N_f)}(\zeta)
                                      &-\zeta^{j-1}\sinh^{(j-1)} (\zeta)\\
\om^{j-1}\cosh^{(j-1)} (\om)          &\om^{N_f+1} \cosh^{(N_f)}(\om)
                                      &-\om^{j-1}\sinh^{(j-1)} (\om)\\

\end{array}\right)
\right. \nn\\
&&+
\det\left(
\begin{array}{lll}
{\bf 0}         & 0                                   & -\matS(\{\mu_i\})\\
\matC(\{\mu_i\})&\zeta\mu_i^{N_f}\cosh^{(N_f)}(\mu_i) &\ \ \matS(\{\mu_i\})\\
\zeta^{j-1}\cosh^{(j-1)} (\zeta)      &\om\zeta^{N_f} \cosh^{(N_f)}(\zeta)
                                      &-\zeta^{j-1}\sinh^{(j-1)} (\zeta)\\

\om^{j-1}\cosh^{(j-1)} (\om)          &\zeta\om^{N_f} \cosh^{(N_f)}(\om)
                                      &-\om^{j-1}\sinh^{(j-1)} (\om)

\end{array}\right)
\nn\\
&&-
\det\left(
\begin{array}{lll}
{\bf 0}          & 0                                  & -\matS(\{\mu_i\})\\
\matC(\{\mu_i\}) & \om\mu_i^{N_f}\cosh^{(N_f)}(\mu_i)&\ \  \matS(\{\mu_i\})\\
\zeta^{j-1}\cosh^{(j-1)} (\zeta)      &-\zeta^{N_f+1} \cosh^{(N_f)}(\zeta)
                                      &\zeta^{j-1}\sinh^{(j-1)} (\zeta)\\
\om^{j-1}\cosh^{(j-1)} (\om)          &\om^{N_f+1} \cosh^{(N_f)}(\om)
                                      &-\om^{j-1}\sinh^{(j-1)} (\om)

\end{array}\!\right)
\nn\\
&&- \left.
\det\left(
\begin{array}{lll}
{\bf 0}         &\ \  0                               & -\matS(\{\mu_i\})\\
\matC(\{\mu_i\})&-\zeta\mu_i^{N_f}\cosh^{(N_f)}(\mu_i)&\ \ \matS(\{\mu_i\})\\
\zeta^{j-1}\cosh^{(j-1)} (\zeta)     &\om\zeta^{N_f} \cosh^{(N_f)}(\zeta)
                                     &\zeta^{j-1}\sinh^{(j-1)} (\zeta)\\
\om^{j-1}\cosh^{(j-1)} (\om)         &-\zeta\om^{N_f}\cosh^{(N_f)}(\om)
                                     &-\om^{j-1}\sinh^{(j-1)} (\om)
\end{array}\!\right)
\!\right]\!. \nn
\eeqn
In the next step we can again add the first and third determinant
as well as the second and fourth determinant,
after multiplying in the fourth determinant the last column and
last but one row with $(-1)$. We obtain
\beqn
(\om+\zeta)\det {\matD}(\{\mu_i\},\zeta,\om) &-&
(\om-\zeta)\det {\matD}(\{\mu_i\},-\zeta,\om)  \ =\   \label{det2step1}   \\
\ =\  2^{N_f+1}&&\left[
\det\left(
\begin{array}{lll}
{\bf 0}                    & 0                 & -\matS(\{\mu_i\})\\
\matC(\{\mu_i\})           & 0                 &\ \ \matS(\{\mu_i\})\\
 0                         &\zeta^{N_f+1}\cosh^{(N_f)}(\zeta)
                           &-\zeta^{j-1}\sinh^{(j-1)} (\zeta)\\
\om^{j-1}\cosh^{(j-1)}(\om)&\om^{N_f+1} \cosh^{(N_f)}(\om)
                           &-\om^{j-1}  \sinh^{(j-1)} (\om)

\end{array}\!\right)\right.
\nn\\
+&&
\det\left(
\begin{array}{lll}
{\bf 0}                    & 0 & -\matS(\{\mu_i\})\\
\matC(\{\mu_i\})           &\om\mu_i^{N_f}\cosh^{(N_f)}(\mu_i)
                           &\ \ \matS(\{\mu_i\}\\
   0                       & 0 &-\zeta^{j-1}\sinh^{(j-1)}(\zeta)\\
\om^{j-1}\cosh^{(j-1)}(\om)& 0 &-\om^{j-1}\sinh^{(j-1)} (\om)
\end{array}\right)\nn\\
+&&\left.
\det\left(
\begin{array}{lll}
{\bf 0}                        & 0 & -\matS(\{\mu_i\})\\
\matC(\{\mu_i\})               &\zeta\mu_i^{N_f}\cosh^{(N_f)}(\mu_i)
                               &\ \ \matS(\{\mu_i\})  \\
\zeta^{j-1}\cosh^{(j-1)}(\zeta)& 0 & 0 \\
\om^{j-1}\cosh^{(j-1)} (\om)   &\zeta\om^{N_f} \cosh^{(N_f)}(\om)
                               &-\om^{j-1}\sinh^{(j-1)} (\om)
\end{array}\right)
\!\right], \nn
\eeqn
where we have split the determinant resulting out of the first
addition into two, with respect to the last column,
and taken out a factor of 2. The first determinant in the last
equation is already part of the desired result. In the $(N_f+1)$-st column we
can write $\cosh^{(N_f)}=\sinh^{(N_f+1)}$. If we then permute it to the last
column, the last $N_f+2$ columns only contain
derivatives of $\sinh$. We then Laplace-expand the resulting determinant
into products of $N_f\times N_f$ blocks and
$(N_f+2)\times(N_f+2)$ blocks and obtain
\beq
(-)^{N_f}2^{N_f+1} \det\matC(\{\mu_i\})
\det\left(
\begin{array}{ll}
\matS(\{\mu_i\})                & 0\\
\zeta^{j-1}\sinh^{(j-1)}(\zeta) &\zeta^{N_f+1}\sinh^{(N_f+1)}(\zeta)\\
  \om^{j-1}\sinh^{(j-1)}( \om)  &\om^{N_f+1}  \sinh^{(N_f+1)}(\om)
\end{array}
\right)
\label{det2piece1}
\eeq
after re-arranging all signs. What remains to be shown is that the two
remaining determinants in eq. (\ref{det2step1}) give rise to the same
determinant as in eq. (\ref{det2piece1}), with the last column vector
reading $(\mu_i^{N_f+1}\sinh^{(N_f+1)}(\mu_i),0,0)$ instead.

\noi
We find again that in the third determinant in eq. (\ref{det2step1}), the
last term in the $(N_f+1)$-st column can be set to zero following the
same argument as before.
Next, in both the second and third determinant
in  eq. (\ref{det2step1}) we take the common factors $\zeta$ and
$\om$ out of the $(N_f+1)$-st column, respectively, and multiply them into the
last but one row. The two determinants are then equal up to precisely
the last but one row and can thus be added,
\beq
2^{N_f+1} \det\left(
\begin{array}{lll}
{\bf 0}                      & 0 & -\matS(\{\mu_i\})\\
\matC(\{\mu_i\})             & \mu_i^{N_f}\cosh^{(N_f)}(\mu_i)
                             &\ \ \matS(\{\mu_i\})\\
\zeta^{j}\cosh^{(j-1)}(\zeta)& 0 & -\om\zeta^{j}\sinh^{(j-1)} (\zeta)\\
\om^{j-1}\cosh^{(j-1)} (\om) & 0 &-\om^{j-1}\sinh^{(j-1)} (\om)
\end{array}\right).
\eeq
Along the lines of Lemma 1 it is again straightforward to see,
that the factor $\om$ multiplying the right side of the last but
one row can be multiplied into the left side of the last row
\beq
2^{N_f+1} \det\left(
\begin{array}{lll}
{\bf 0}                      & 0 & -\matS(\{\mu_i\})\\
\matC(\{\mu_i\})             & \mu_i^{N_f}\cosh^{(N_f)}(\mu_i)
                             &\ \ \matS(\{\mu_i\})\\
\zeta^{j}\cosh^{(j-1)}(\zeta)& 0 &-\zeta^{j-1}\sinh^{(j-1)} (\zeta)\\
\om^{j}\cosh^{(j-1)} (\om)   & 0 &-\om^{j-1}\sinh^{(j-1)} (\om)
\end{array}\right).
\label{det2step2}
\eeq
This remains to be shown to be equal to the following product:
\beq
(-)^{N_f}2^{N_f+1} \det\matC(\{\mu_i\})
\det\left(
\begin{array}{ll}
\matS(\{\mu_i\})                &\mu_i^{N_f+1}\sinh^{(N_f+1)}(\mu_i)\\
\zeta^{j-1}\sinh^{(j-1)}(\zeta) & 0 \\
  \om^{j-1}\sinh^{(j-1)}( \om)  & 0
\end{array}
\right),
\label{det2piece2}
\eeq
which then adds up to eq. (\ref{det2piece1}) to give
eq. (\ref{La2}). To proceed,
we eliminate in  eq. (\ref{det2step2}) the lower block
$\matS$ by adding the upper one.
Next, we expand the determinant
into products of $(N_f+2)\times(N_f+2)$ blocks
and $N_f\times N_f$  blocks. This time it will be the last blocks,
that give us the prefactor $\det\matC((\{\mu_i\})$
\beqn
&&-2^{N_f+1}
\left\{
\det\left(
\begin{array}{lll}
\matC(\{\mu_i\})             &\mu_i^{N_f}\cosh^{(N_f)}(\mu_i)& 0\\
\zeta^{j}\cosh^{(j-1)}(\zeta)& 0                             & \sinh(\zeta)\\
\om^{j}\cosh^{(j-1)}(\om)    & 0                             & \sinh(\om)\\
\end{array}\right)
\det\left(\mu_i^j\sinh^{(j)}(\mu_i)\right)
\right.
\label{det2step3}\\
&&\ \ \ \ \ \ -\sum_{k=1}^{N_f}(-1)^k \left[
\det\left(
\begin{array}{lll}
 0                           & 0                             &\sinh(\mu_k)\\
\matC(\{\mu_i\})             &\mu_i^{N_f}\cosh^{(N_f)}(\mu_i)& 0\\
\zeta^{j}\sinh^{(j-1)}(\zeta)& 0                             &\sinh(\zeta)
\end{array}\right)
\det\left(
\begin{array}{l}
\mu_i^j\sinh^{(j)}(\mu_i)\\
\ldots \mbox{no}\ k\ldots\\
\om^{j}\sinh^{(j)}(\om)
\end{array}\right)
\right.
\nn\\
&&\ \ \ \ \ \ \ \ \ \ \ \ \ \ \ \ \ \ -\ \left.\left.
\det\left(
\begin{array}{lll}
 0                       & 0                             &\sinh(\mu_k)\\
\matC(\{\mu_i\})         &\mu_i^{N_f}\cosh^{(N_f)}(\mu_i)& 0\\
\om^{j}\sinh^{(j-1)}(\om)& 0                             &\sinh(\om)
\end{array}\right)
\det\left(
\begin{array}{l}
\mu_i^j\sinh^{(j)}(\mu_i)\\
\ldots \mbox{no}\ k\ldots\\
\zeta^{j}\sinh^{(j)}(\zeta)
\end{array}\right)
\!\right]\!\right\}. \nn
\eeqn
The first product gives part of eq. (\ref{det2piece2})
by taking $\prod_i \mu_i$ out of the second determinant and
multiplying the first $N_f$ rows of the first determinant, upon
rewriting $\cosh^{(j)}=\sinh^{(j-1)}$,
\beq
(-)^{N_f}2^{N_f+1}
\det\left(
\begin{array}{lll}
0           &\mu_i^{j}\sinh^{(j)}(\mu_i)&\mu_i^{N_f+1}\sinh^{(N_f+1)}(\mu_i)\\
\sinh(\zeta)&\zeta^{j}\sinh^{(j)}(\zeta)& 0 \\
\sinh( \om) &  \om^{j}\sinh^{(j)}( \om) & 0
\end{array}
\right) \det\matC(\{\mu_i\})\ .
\label{det2piece2a}
\eeq
Next, we tackle the sum over $k$ in eq. (\ref{det2step3}).
Along the lines of Lemma 1 we expand the first determinant
into products of  $N_f\times N_f$ blocks and
$2\times2$ blocks,
\beqn
&&2^{N_f+1}\sum_{k,l=1}^{N_f}(-1)^{k+l}
\left[
\det\left(
\begin{array}{l}
\matC(\{\mu_{i\neq l}\})\\
\zeta^{j}\cosh^{(j-1)}(\zeta)
\end{array}\right)
\sinh(\mu_k)\ \mu_l^{N_f}\!\cosh^{(N_f)}(\mu_l)
\det\left(
\begin{array}{l}
\mu_i^j\cosh^{(j-1)}(\mu_i)\\
\ldots \mbox{no}\ k\ldots\\
\om^{j}\cosh^{(j-1)}(\om)
\end{array}\right)
\right.\nn\\
&&\left.
\ \ \ \ \ \ \ \ \ \ \ \ \ \ \ \ \ \ \ \ \ \ \ -
\det\left(
\begin{array}{l}
\matC(\{\mu_{i\neq l}\})\\
\om^{j}\cosh^{(j-1)}(\om)
\end{array}\right)
\sinh(\mu_k)\ \mu_l^{N_f}\!\cosh^{(N_f)}(\mu_l)
\det\left(
\begin{array}{l}
\mu_i^j\cosh^{(j-1)}(\mu_i)\\
\ldots \mbox{no}\ k\ldots\\
\zeta^{j}\cosh^{(j-1)}(\zeta)
\end{array}\right)
\right]\nn\\
&&=\ 2^{N_f+1}\mbox{\Huge[}
\sum_{k,l=1}^{N_f}(-1)^{k+l}
\zeta\om\!\prod_{j\neq k,l}\mu_j\
\sinh(\mu_k)\ \mu_l^{N_f+1}\sinh^{(N_f+1)}(\mu_l)
\nn\\
&&\ \ \ \ \ \ \ \ \ \ \times\left(
\det\matC(\{\mu_{i\neq k}\},\om) \det\matC(\{\mu_{i\neq l}\},\zeta)
\ -\
\det\matC(\{\mu_{i\neq k}\},\zeta) \det\matC(\{\mu_{i\neq l}\},\om)
\right)\mbox{\Huge]}\nn\\
&&=\ 2^{N_f+1}\left[
\sum_{k>l}^{N_f}(-1)^{k+l}
\zeta\om\!\prod_{j\neq k,l}\mu_j\
\sinh(\mu_k)\ \mu_l^{N_f+1}\!\sinh^{(N_f+1)}(\mu_l)
\det\matC(\{\mu_{i}\}) \det\matC(\{\mu_{i\neq l,k}\},\zeta,\om)\right.\nn\\
&&\ \ \ \ \ \ \ \ \ \ - \left.
\sum_{k<l}^{N_f}(-1)^{k+l}
\zeta\om\!\prod_{j\neq k,l}\mu_j\
\sinh(\mu_k)\ \mu_l^{N_f+1}\!\sinh^{(N_f+1)}(\mu_l)
\det\matC(\{\mu_{i}\}) \det\matC(\{\mu_{i\neq k,l}\},\zeta,\om)\right]\nn\\
&&=\ 2^{N_f+1}\det\matC(\{\mu_{i}\}) \nn\\
&&\ \ \ \ \times\sum_{k>l}^{N_f}(-1)^{k+l}
\zeta\om\!\prod_{j\neq k,l}\mu_j
\det\left(
\begin{array}{ll}
\sinh(\mu_k)&\mu_k^{N_f+1}\!\sinh^{(N_f+1)}(\mu_k)\\
\sinh(\mu_l)&\mu_l^{N_f+1}\!\sinh^{(N_f+1)}(\mu_l)
\end{array}\right)
\det\matC(\{\mu_{i\neq k,l}\},\zeta,\om)
\nn\\
&&=\ (-1)^{N_f}2^{N_f+1}
\det\matC(\{\mu_i\})\
\det\left(
\begin{array}{lll}
\sinh(\mu_i)&\mu_i^{j}\sinh^{(j)}(\mu_i)
                                       &\mu_i^{N_f+1}\sinh^{(N_f+1)}(\mu_i)\\
\ \ 0       &\zeta^{j}\sinh^{(j)}(\zeta)& \ \ 0\\
\ \ 0       &\om^{j}\sinh^{(j)} (\om)     & \ \ 0
\end{array}\right) .
\label{det2piece2b}
\eeqn
Here, we have performed the same steps as in the respective part of the
proof of Lemma 1.
Thus eqs. (\ref{det2piece2b}) and (\ref{det2piece2a}) sum up to
eq. (\ref{det2piece2}), which together with
eq. (\ref{det2piece1}) leads to the right hand side of eq. (\ref{La2}).

\setcounter{equation}{0}
\section{Relations between QCD$_3$ and QCD$_4$ from Random
Matrix Theory} \label{B}
\vspace*{0.3cm}

\noi
In this subsection we show that the relations of
Theorem III also can be derived from an entirely different point of view,
based on the Random Matrix Theory representations of
the QCD$_3$ and QCD$_4$ finite-volume partition functions.

\noi
Using the results of refs. \cite{VZ,SV}, we write the
partition functions (\ref{ZQCD3even}) and  (\ref{ZchUE}) as the
large-${\cal N}$
limits of the matrix integrals
\beq
\tilde{\cal Z}_{{\rm QCD}_3}^{(2N_{f})}(\{m_i\})
~=~ \int\! d\phi \prod_{f=1}^{2N_{f}}{\det}\left(\phi + im_f\right)~
\exp\left[-{\cal N}\mbox{tr}\, V(\phi^2)\right] ~,
\label{ZRMTUE}
\eeq
for QCD$_3$, and
\beq
\tilde{\cal Z}_{\nu}^{(N_{f})}(\{m_i\})
~=~ \int\! dW \prod_{f=1}^{N_{f}}{\det}\left(i\bar{\phi} + m_f\right)~
\exp\left[-\frac{{\cal N}}{2} \mbox{tr}\, V(\bar{\phi}^2)\right] ~,
\label{ZRMTchUE}
\eeq
for QCD$_4$.
Here $\phi$ is a Hermitian $2{\cal N}\times 2{\cal N}$ matrix and $\bar \phi$
\beq
\bar{\phi} ~=~ \left( \begin{array}{cc}
              0 & W^{\dagger} \\
              W & 0
              \end{array}
      \right) ~,
\eeq
with $W$ an arbitrary complex matrix  of size $({\cal N}+\nu)\times {\cal N}$.
In both cases the Random Matrix Theory potentials
$V(\phi^2)$ are not constrained
beyond giving convergent integrals with a non-vanishing spectral density
at the origin, $\rho(0) \neq 0$ \cite{ADMN}. In eq. (\ref{ZRMTUE}) the
masses are taken to be pairwise grouped with opposite signs exactly
as in the integral (\ref{ZQCD3even}).
The limit ${\cal N}\to \infty$ is
taken in the two Random Matrix Theory partition functions
so that $\mu_i \equiv m_i\pi\rho(0){\cal N}$ and
$\mu_i \equiv m_i\pi\rho(0)2{\cal N}$,
respectively, are kept fixed. In this limit the partition functions
${\cal Z}_{{\rm QCD}_3}^{(2N_f)}$ and
$\tilde{\cal Z}_{{\rm QCD}_3}^{(2N_{f})}$ become
proportional to each other, with
a $\mu_i$-independent proportionality constant
\cite{VZ}. Similarly, also the partition functions
${\cal Z}^{(N_f)}_{\nu}$ and $\tilde{\cal Z}_{\nu}^{(N_{f})}$ become
proportional \cite{SV}.
Below we derive
relations between the unitary matrix integrals (\ref{ZQCD3even})
and (\ref{ZchUE}) from
their Random Matrix representation (\ref{ZRMTUE}) and
(\ref{ZRMTchUE}), respectively.
The key ingredient is to use the standard
orthogonal polynomial technique of Random Matrix Theory, and next
rewrite all expressions in terms of the partition functions themselves
\cite{D,AD1}. Additional details can be found
in ref. \cite{AD1}.

\noi
We begin by expressing the integrals
(\ref{ZRMTUE}) and (\ref{ZRMTchUE}) in terms of
eigenvalue representations. This permits us to consider
non-integer values of $\nu$ through analytical continuation.
The starting point is a relation between the orthogonal polynomials of
chUE (QCD$_4$) and those of the UE (QCD$_3$):
\beq
P_{l,\ {\rm chUE}}^{(N_f,\ \nu=-1/2)} (\{m_i\};z^2) \ =\
P_{2l,\ {\rm UE}}^{(2N_f)}\ (\{m_i\};z) \ .
\label{OP}
\eeq
This simple identity is easily derived from the definition of the
two Random Matrix Theories involved \cite{ADMN2}.
Since in ref. \cite{ADMN2} no explicit use was made of the measure,
eq. (\r{OP}) also holds in the massive case.
On the right hand side the same $N_f$ masses appear, but doubled into pairs
with opposite sign.

\noi
Instead of using directly the orthogonal polynomials $P_l(\{m_i\};\lambda)$,
it is convenient to work with the ``wave functions'',
\beq
\Psi_l(\{m_i\};\lambda) ~\equiv~ \sqrt{\omega(\{m_i\},\lambda)}
P_l(\{m_i\};\lambda) ~,
\eeq
where $\omega(\{m_i\};\lambda)$ is the measure function (so that the wave
functions $\Psi_l(\{m_i\};\lambda)$ are orthogonal with respect to a weight
of unity). It follows from eq. (3.11) of ref. \cite{ADMN2} that one can use
the wave functions of the UE inside the Christoffel-Darboux identity
for the chUE kernel. We then find the following identity \cite{AD2}:
\newpage
\beqn
K_{{\cal N},\ {\rm chUE}}^{(N_f,\ \nu=-1/2)} (z^2,w^2) &=&
\frac{c_{2{\cal N}}^{(2N_f)}}{z^2-w^2}
\left( \Psi_{2{\cal N},\ {\rm UE}}^{(2N_f)}(z)\ w\ \Psi_{2{\cal N}-1,\
{\rm  UE}}^{(2N_f)}(w)
\ -\ z\ \Psi_{2{\cal N}-1,\ {\rm  UE}}^{(2N_f)}(z)\Psi_{2{\cal N},\
{\rm  UE}}^{(2N_f)}(w)
\right) \nn\\
&=&\frac{1}{2}\frac{c_{2{\cal N}}^{(2N_f)}}{(z-w)}
\left( \Psi_{2{\cal N},\ {\rm UE}}^{(2N_f)}(z)\Psi_{2{\cal N}-1,\
{\rm UE}}^{(2N_f)}(w)
\ -\ \Psi_{2{\cal N}-1,\ {\rm UE}}^{(2N_f)}(z)\Psi_{2{\cal N},\
{\rm  UE}}^{(2N_f)}(w)
\right)  \nn\\
&&-\ \frac{1}{2}\frac{c_{2{\cal N}}^{(2N_f)}}{(z+w)}
\left( \Psi_{2{\cal N},\ {\rm UE}}^{(2N_f)}(z)\Psi_{2{\cal N}-1,\
{\rm UE}}^{(2N_f)}(w)
\ + \ \Psi_{2{\cal N}-1,\ {\rm UE}}^{(2N_f)}(z)\Psi_{2{\cal N},\
{\rm  UE}}^{(2N_f)}(w)
\right)  \nn\\
&=&
K_{2{\cal N},\ {\rm UE}}^{(2N_f)} (z,w) \ +\
K_{2{\cal N},\ {\rm  UE}}^{(2N_f)} (-z,w) \ ,
\label{kernel}
\eeqn
where for clarity we have not explicitly indicated the mass dependence of the
wave functions.
The coefficients $c_{2{\cal N}}^{(2N_f)}$ are defined in ref. \cite{ADMN2}.
We have also used that the parity of the polynomials $P_l(\la)$ for the UE
is equal to $(-1)^l$.
The different kernels can all be expressed in terms the corresponding
finite volume partition functions \cite{D,AD1}, as we will see below.
In fact, the relation (\ref{kernel}) simply expresses Lemma I of
Appendix A, divided on each side by the result of Theorem I.

\noi
In order to arrive at a more general result, we will also
derive an equation involving ${\cal Z}_{\nu=+1/2}^{(N_f)}(\{\mu\})$.
To do so we start with the kernel $K_{{\cal N},\ {\rm chUE}}^{(N_f,\
  \nu=+1/2)} (z^2,w^2)$ with $\nu=+1/2$ in contrast to
eq. (\r{kernel}). Here, we will make use of the following relation
among the wavefunctions of the UE:
\beq
 \Psi_{2l,\ {\rm UE}}^{(2N_f,2\alpha +2)}(z)~=~ \sign(z)\Psi_{2l+1,\
  {\rm  UE}}^{(2N_f,2\alpha)}(z) ~.
\label{psirelation}
\eeq
This relation was shown in \cite{ADMN2} for any number of massless
flavors $2\alpha$, but it can  easily be generalized to include any
additional number of  $2N_f$  (massless or massive) flavors as well,
since they are merely spectators.
The sign function takes into account that the left hand side is an even
function. Employing again the relation eq. (\r{OP}) we obtain
\beqn
K_{{\cal N},\ {\rm chUE}}^{(N_f,\ \nu=+1/2)} (z^2,w^2) &=&
\frac{c_{2{\cal N}}^{(2N_f+2)}c_{2{\cal N}-1}^{(2N_f+2)}}{z^2-w^2}\!
\left( \Psi_{2{\cal N},\ {\rm  UE}}^{(2N_f+2)}(z)\Psi_{2{\cal N}-2,\:
{\rm UE}}^{(2N_f+2)}(w)
- \Psi_{2{\cal N},\ {\rm UE}}^{(2N_f+2)}(w)\Psi_{2{\cal N}-2,\:
{\rm  UE}}^{(2N_f+2)}(z)
\right)
\nn\\
&=& \frac{c_{2{\cal N}}^{(2N_f+2)}c_{2{\cal
      N}-1}^{(2N_f+2)}}{z^2-w^2}\
\sign(z)\sign(w)\nn\\
&&\left( \Psi_{2{\cal N}+1,\ {\rm UE}}^{(2N_f)}(z)\Psi_{2{\cal N}-1,\
{\rm UE}}^{(2N_f)}(w)
\ -\  \Psi_{2{\cal N}+1,\ {\rm UE}}^{(2N_f)}(w)\Psi_{2{\cal N}-1,\
{\rm UE}}^{(2N_f)}(z)
\right)\nn\\
&=& \frac{c_{2{\cal N}}^{(2N_f+2)}c_{2{\cal N}-1}^{(2N_f+2)}}
{c_{2{\cal N}}^{(2N_f)}}
\frac{1}{z^2-w^2}\ \sign(z)\sign(w)\nn\\
&&\left( \Psi_{2{\cal N}+1,\ {\rm UE}}^{(2N_f)}(z)\ w\
\Psi_{2{\cal N},\ {\rm UE}}^{(2N_f)}(w)
\ -\Psi_{2{\cal N}+1,\ {\rm UE}}^{(2N_f)}(w)\ z\ \Psi_{2{\cal N},\
{\rm UE}}^{(2N_f)}(z)
\right) ~.
\eeqn
In the first step we have used the relation (\r{psirelation}) and in
the second step the recursion relation
$\la P_l(\la)=c_{l+1}P_{l+1}(\la)+c_lP_{l-1}(\la)$.
With the same rewriting as in
eq. (\r{kernel}) we arrive at
\beqn
K_{{\cal N},\ {\rm chUE}}^{(N_f,\ \nu=+1/2)} (z^2,w^2)
&=&
\frac{c_{2{\cal N}}^{(2N_f+2)}c_{2{\cal N}-1}^{(2N_f+2)}}
{c_{2{\cal N}}^{(2N_f)}}\ \sign(z)\sign(w)
\label{kernel2}\\
&&\frac{1}{2}\left[\frac{1}{(z-w)}
\left( \Psi_{2{\cal N}+1,\ {\rm UE}}^{(2N_f)}(z)\Psi_{2{\cal N},\
{\rm UE}}^{(2N_f)}(w)
\ - \ \Psi_{2{\cal N},\ {\rm UE}}^{(2N_f)}(z)\Psi_{2{\cal N}+1,\
{\rm UE}}^{(2N_f)}(w)
\right)  \right.
\nn\\
&&\ +\ \left. \frac{1}{(z+w)}
\left( - \Psi_{2{\cal N}+1,\ {\rm UE}}^{(2N_f)}(z)\Psi_{2{\cal N},\
{\rm UE}}^{(2N_f)}(w)
\ -\ \Psi_{2{\cal N},\ {\rm UE}}^{(2N_f)}(z)\Psi_{2{\cal N}+1,\
 {\rm UE}}^{(2N_f)}(w)\right) \right]\nn\\
&=&  \sign(z)\sign(w)
\frac{c_{2{\cal N}}^{(2N_f+2)}c_{2{\cal N}-1}^{(2N_f+2)}}
{c_{2{\cal N}}^{(2N_f)}c_{2{\cal N}+1}^{(2N_f)}}
\left(
K_{2{\cal N}+1,\ {\rm UE}}^{(2N_f)} (z,w) -
K_{2{\cal N}+1,\ {\rm UE}}^{(2N_f)} (-z,w)
\right).\nn
\eeqn
In the large$-{\cal N}$ limit the mass dependence of the recursion
coefficients
$c_{2{\cal N}}^{(2N_f)}$ becomes subleading and these coefficients
simply cancel out.
Again, this relation has a direct interpretation in terms of finite
volume partition functions: it is Lemma II of Appendix B divided on
each side by the relation of Theorem I.

\noi
Adding and subtracting eqs. (\r{kernel}) and (\r{kernel2})
we finally arrive at
\beq
K_{\rm UE}^{(2N_{f})}(\{\mu_i\},\pm\zeta,\omega) ~=~ \frac12 \left(
K_{\rm chUE}^{(N_{f},\nu=-1/2)}(\{\mu_i\},\zeta^2,\omega^2) ~\pm~
K_{\rm chUE}^{(N_{f},\nu=+1/2)}(\{\mu_i\},\zeta^2,\omega^2) \right)
\label{kernels1}
\eeq
in the microscopic large-${\cal N}$ limit, for $\zeta,\omega\geq0$. One of
these (the one with ``+'' signs) was previously known only in the
massless case, where it can readily be derived from the generalized
Laguerre ensemble \cite{H}.

\noi
We now make the following observation. From the kernel relation
(\ref{kernels1}) it follows that all spectral correlation functions
of the two Random Matrix Theories (\ref{ZRMTUE}) and (\ref{ZRMTchUE}) are
related in a fairly simple way. If one believes that these relations extend
to the microscopic spectral correlators of the Dirac operators in QCD$_3$ and
QCD$_4$, then one obtains all spectral correlation functions
$\rho_{QCD_{3}}^{(2N_{f})}(\zeta_1,\ldots,\zeta_k;\{\mu_i\})$
in QCD$_3$ from the kernel relevant for QCD$_4$.
The most obvious relation that follows from
(\ref{kernels1}) is the one for the microscopic spectral densities,
\beq
\rho_{{\rm QCD}_{3}}^{(2N_{f})}(\zeta;\{\mu_i\}) ~=~
\frac{1}{2}\left[\rho_{{\rm QCD}_4}^{(N_{f},\nu=+1/2)}(\zeta;\{\mu_i\}) +
\rho_{{\rm QCD}_4}^{(N_{f},\nu=-1/2)}(\zeta;\{\mu_i\})\right].
\label{rho34evenRMT}
\eeq
This relation is here derived within the context of Random Matrix Theory.
As shown in this paper, it also follows
directly from the field theory formulation, through the replica
method.

\noi
We note that
based on the exact relationship of two-point correlation functions
to the Random Matrix Theory
partition functions \cite{D,AD1}
(which in turn are proportional to the finite-volume field
theory partition functions), eq. (\ref{kernels1}) implies
the following identity:
\beq
\frac{{\cal Z}^{(2N_{f}+2)}_{{\rm QCD}_3}(\{\mu_i\},\pm\zeta,\omega)}
{{\cal Z}^{(2N_{f})}_{{\rm QCD}_3}(\{\mu_i\})} ~=~
\pi(\zeta\omega)^{1/2}
\left(
\frac{{\cal Z}^{(N_{f}+2)}_{\nu=-1/2}(\{\mu_i\},\zeta,\omega)}
{{\cal Z}^{(N_{f})}_{\nu=-1/2}(\{\mu_i\})} \mp
\frac{{\cal Z}^{(N_{f}+2)}_{\nu=+1/2}(\{\mu_i\},\zeta,\omega)}
{{\cal Z}^{(N_{f})}_{\nu=+1/2}(\{\mu_i\})}
\right)~.
\label{Zrelation}
\eeq
This particular relation is simply the identity of Theorem III
divided on each side by the identity of Theorem I, both of which
were proven in section 4.1. The right hand side involves a
sum or difference of two different ratios of partition functions that
cannot easily be combined. The identity as it stands is therefore
not suitable for
deriving, by means of the replica method, identities among spectral
correlators of the Dirac operators in
the two theories. We note, however, that in the limit,
$\zeta,\omega \to 0$ it is evidently consistent with the general identity
(\ref{factoreven}). In this limit the second term on the
right hand side of (\ref{Zrelation}) vanishes. Although the first
term is also multiplied by $(\zeta\omega)^{1/2}$, the product of the
two remains finite. Indeed, invoking {\em flavor-topology duality}
\cite{Jac}, we can express the product in terms of a partition function
with two massless particle less, at the cost of increasing $\nu$,
\beq
\frac{{\cal Z}^{(2N_{f}+2)}_{{\rm QCD}_3}(\{\mu_i\},0,0)}
{{\cal Z}^{(2N_{f})}_{{\rm QCD}_3}(\{\mu_i\})} ~=~
2\prod_{j=1}^{N_f}\frac{1}{\mu_j^2}
\frac{{\cal Z}^{(N_{f})}_{\nu=+3/2}(\{\mu_i\})}
{{\cal Z}^{(N_{f})}_{\nu=-1/2}(\{\mu_i\})} ~.
\label{Z34even}
\eeq

\noi
By applying flavor-topology duality,
the relation (\ref{Z34even}) also follows from the more
general identity (\ref{factoreven}) given in Theorem I alone.
To see this, we take the ratio of
eq. (\ref{factoreven}) for $N_f+1$ and $N_f$ and then send the extra
mass to zero.
The factors
${\cal Z}_{+1/2}^{(N_f)}(\{\mu_i\})$ cancel after the flavor-topology
shift (\ref{FTfactornu})                                      
on the partition functions with $\nu=-1/2$ and $\nu=+1/2$
and we arrive at eq. (\ref{Z34even}).

\setcounter{equation}{0}
\section{Partition Function and
Spectral Sum Rules to Sixth Order}
\vspace*{0.3cm}

\noi\indent
In this Appendix we give the terms of order $\mu^5$ and $\mu^6$
in the small expansion of the QCD$_3$ partition function ${\cal Z}_q^{(N)}$,
The terms up to order $\mu^4$ were already given in
eq. (\ref{m6}). For the terms of $O(\mu^4)$ we find
\beqn
{\cal Z}_q^{(N)}\vert_{\mu^5}  &=& \,
 -\frac{2q(N^2-q^2)(20-3N^2+7q^2)}{5N(N^2-1)(N^2-4)(N^2-9)(N^2-16)}
{\rm Tr}M^5 \label{z5}\\
 &-& \frac{q(N^2-q^2)\left[N^4+N^2(4-5q^2)+24
(q^2-4)\right]}{4N^2(N^2-1)(N^2-4)(N^2-9)(N^2-16)}{\rm Tr}M^4 {\rm Tr}M
 \nn\\
&+& \frac{q(N^2-q^2)(N^2q^2+12q^2-N^4+12N^2-48)}
{3N^2(N^2-1)(N^2-4)(N^2-9)(N^2-16)}{\rm Tr}M^3{\rm Tr}M^2 \nn\\
&+&
\frac{q(N^2-q^2)(4-q^2)}{3N(N^2-1)(N^2-4)(N^2-16)}{\rm Tr}M^3({\rm Tr}M)^2
\nn\\
&+&
\frac{q(N^2-q^2)(N^4-N^2q^2-18N^2+2q^2+88)}
{8N(N^2-1)(N^2-4)(N^2-9)(N^2-16)}
{\rm Tr}M ({\rm Tr}M^2)^2 \nn\\
&+& \frac{
q(N^2-q^2)(N^4q^2-14N^2q^2+24q^2-3N^4+40N^2-96)}
{12N^2(N^2-1)(N^2-4)(N^2-9)(N^2-16)}{\rm Tr}M^2({\rm Tr}M)^3\nn\\
&+&\frac{q\left[ N^4(q^4-10q^2+15)-20N^2(q^4-9q^2+11)+
6(13q^4-100q^2+96)\right]}{120N(N^2-1)(N^2-4)(N^2-9)(N^2-16)}
({\rm Tr}M)^5. \nn
\eeqn
At order $\mu^6$ we find
\beqn
{\cal Z}_q^{(N)}\vert_{\mu^6} \, &=& \,
\sum_{l=0}^5A_l {\rm Tr}M^{6-l}({\rm Tr}M)^l
+\sum_{k=1}^2B_k {\rm Tr}M^{6-2k}({\rm Tr}M^2)^k \nn\\
&+& \sum_{i=1}^3C_i\left({\rm Tr}M^i{\rm Tr}M^{3-i}\right)^2 +D \left(
{\rm Tr}M {\rm Tr}M^2 {\rm Tr}M^3 \right).
\label{ABCD}
\eeqn
where
\beqn
A_0&=&\frac{(N^2-q^2)\left[(N^2-4)(N^2-16)+140q^2 - 14N^2q^2 + 21q^4\right]}
{3N(N^2-1)(N^2-4)(N^2-9)(N^2-16)(N^2-25)},\nn\\
A_1&=& \frac{2(N^2-q^2)\left[-5(N^2-4)(N^2-16)+q^2(-200-25N^2
+3N^4+70q^2-7N^2q^2)\right]}{5N^2(N^2-1)(N^2-4)(N^2-9)(N^2-16)(N^2-25)}
,\nn\\
A_2&=&\, \frac{(1-q^2)(N^2-q^2)\left[ (N^2-4)(N^2+23) +13q^2-5N^2q^2\right]}
{8N(N^2-1)^2(N^2-4)(N^2-9)(N^2-25)},\nn\\
A_3&=&\, \frac{-(N^2-q^2)}{18N^2(N^2-1)^2(N^2-4)(N^2-9)(N^2-16)(N^2-25)}
\bigg[(N^2-4)(N^2-16)(9N^2+15)\nn\\
& &\,  -\, 1541N^2q^2
+356N^4q^2-15N^6q^2-60q^4+229N^2q^4-51N^4q^4+2N^6q^4\bigg],
\nn\\
A_4&=& \frac{(N^2-q^2)
\left[3(N^2-4)(N^2-16)+q^2(-280+136N^2-6N^4+38q^2-24N^2q^2+N^4q^2)
\right]}{48N(N^2-1)^2(N^2-4)(N^2-16)(N^2-25)},\nn\\
A_5&=& \left[720N^2(N^2-1)^2(N^2-4)(N^2-9)(N^2-16)(N^2-25)\right]^{-1}
\!\bigg[15N^2(4-N^2)(N^2-9)(N^2-16) \nn\\
&&+ 5760q^2 -29416N^2q^2 + 13756N^4q^2
- 1545N^6q^2 + 45N^8q^2 - 2400q^4+14290N^2q^4 \nn\\
&&- 5850N^4 q^4 + 575N^6q^4 -
    15N^8q^4 + 240q^6 - 1258N^2q^6 + 458N^4q^6 - 41N^6q^6 + N^8q^6 \bigg],
\nn\\
B_1&=&
\frac{-(N^2-q^2)}{8N^2(N^2-1)^2(N^2-4)(N^2-9)(N^2-16)(N^2-25)}
\bigg[(40-17N^2+N^4)(N^2-4)(N^2-16)\nn\\
& & \, +\, q^2(-800-484N^2+90N^4-6N^6+40q^2+
75N^2q^2+5N^4q^2)\bigg],\nn\\
B_2&=&
\frac{(N^2-q^2)}{48N(N^2-1)^2(N^2-4)(N^2-9)(N^2-16)(N^2-25)}
\bigg[(N^4-27N^2+98)(N^2-4)(N^2-16)\nn\\
& & \, +\, q^2(-3480-164N^2+46N^4-2N^6+358q^2
+N^2q^2+N^4q^2)\bigg],
\nn\\
C_1&=& C_2 = \
\frac{(N^2-q^2)(q^2-1)\left[(N^4-15N^2-10)(N^2-4)+q^2(-10+3N^2-N^4)\right]}
{32N^2(N^2-1)^2(N^2-4)(N^2-9)(N^2-25)},\nn\\
C_3&=&\frac{2(N^2-q^2)}{9N^4(N^2-1)^2(N^2-4)(N^2-9)(N^2-16)(N^2-25)}\bigg[
(15-3N^2)(N^2-4)(N^2-16)\nn\\
& &\, +\, q^2(-600+3N^2-4N^4+N^6+90q^2-29N^2q^2-N^4q^2)\bigg], \nn\\
D&=&
\frac{(N^2-q^2)\left[3(N^2-4)(N^2-16)-q^2
(180-32N^2+2N^4-13q^2-2N^2q^2)\right] }
{6N(N^2-1)^2(N^2-4)(N^2-16)(N^2-25)}.
\label{long}
\eeqn
{}From (\ref{ABCD}) and (\ref{long}) we derive the following
additional spectral sum rules for an {\bf even} number of Dirac eigenvalues:
\beqn
\left\langle\sum_n \frac{1}{\zeta_n^6}\right\rangle
\, =\, 6\, A_0 \quad &,& \quad
\left\langle\sum_n \frac{1}{\zeta_n}
\sum_m\frac{1}{\zeta_m^5} \right\rangle \, = \, -5\, A_1,
\nn\\
\left\langle\left(\sum_n \frac{1}{\zeta_n}\right)^4
\sum_m \frac{1}{\zeta_m^2}\right\rangle
\,=\,  48\, A_4 \quad &,& \quad
\left\langle\sum_n \frac{1}{\zeta_n} \sum_m\frac{1}{\zeta_m^2}
\sum_p\frac{1}{\zeta_p^3} \right\rangle  \, =\,  6\, D, \nn\\
\left\langle\left(\sum_n \frac{1}{\zeta_n}\right)^6
\right\rangle \, = \, - 6!\, A_5  \quad &,& \quad
\left\langle\left(\sum_n \frac{1}{\zeta_n}\right)^2
\sum_m \frac{1}{\zeta_m^4}\right\rangle \, = \,
8\, A_2,
\nn\\
\left\langle\left(\sum_n \frac{1}{\zeta_n}\right)^3\sum_m
\frac{1}{\zeta_m^3}
\right\rangle \, = \, -18\, A_3 \quad &,& \quad
\left\langle\left(\sum_n \frac{1}{\zeta_n}\right)^2
\left(\sum_m \frac{1}{\zeta_m^2}\right)^2\right\rangle \, = \,
-32 \, C_1,
\nn\\
\left\langle\left(\sum_n \frac{1}{\zeta_n^3}\right)^2\right\rangle
\, =\, -18N^2\, C_3 \quad &,& \quad
\left\langle\left(\sum_n \frac{1}{\zeta_n^2}\right)^3\right\rangle
\, =\, 48\, B_2, \nn\\
\left\langle\sum_n \frac{1}{\zeta_n^2}
\sum_m \frac{1}{\zeta_m^4}\right\rangle \, = \,
-8\, B_1. &&
\label{sumrules6}
\eeqn
We remind the reader that $q=0\, , 1$ for even and odd number of  flavors
respectively.

\end{document}